\documentclass[twocolumn]{aastex631}
\usepackage{graphicx,natbib,bm,url,color}
\graphicspath{{./fig/}}

\newcommand{\EQ}{\begin{equation}}
\newcommand{\EN}{\end{equation}}
\newcommand{\EQA}{\begin{eqnarray}}
\newcommand{\ENA}{\end{eqnarray}}

\newcommand{\Eq}[1]{Equation~(\ref{#1})}
\newcommand{\Eqs}[2]{Equations~(\ref{#1}) and~(\ref{#2})}

\newcommand{\Sec}[1]{Section~\ref{#1}}

\newcommand{\Fig}[1]{Figure~\ref{#1}}

\newcommand{\Figs}[2]{Figures~\ref{#1} and \ref{#2}}

\newcommand{\Tab}[1]{Table~\ref{#1}}

\newcommand{\bra}[1]{\langle #1\rangle}
\newcommand{\bbra}[1]{\left\langle #1\right\rangle}

\newcommand{\meanrho}{\overline{\rho}}

{}
{}

{}
{}

{}
{}
{}
{}
{}
{}
{}
{}
{}
{}
{}
{}
{}
{}

{}

{}

{}
{}
{}

\newcommand{\kk}{\bm{k}}

\newcommand{\BB}{\bm{B}}

\newcommand{\JJ}{\bm{J}}
\newcommand{\oo}{\bm{\omega}}

\newcommand{\AAA}{\bm{A}}

\newcommand{\uu}{\bm{u}}

\newcommand{\SSS}{\bm{S}}

\newcommand{\nab}{{\bm{\nabla}}}
\newcommand{\GG}{\mbox{\boldmath $G$} {}}

\newcommand{\SSSS}{\mbox{\boldmath ${\sf S}$} {}}

\newcommand{\ii}{{\rm i}}

\newcommand{\DD}{{\rm D} {}}

\newcommand{\dd}{{\rm d} {}}

\newcommand{\const}{{\rm const}  {}}

\def\la{\mathrel{\mathchoice {\vcenter{\offinterlineskip\halign{\hfil
$\displaystyle##$\hfil\cr<\cr\sim\cr}}}
{\vcenter{\offinterlineskip\halign{\hfil$\textstyle##$\hfil\cr<\cr\sim\cr}}}
{\vcenter{\offinterlineskip\halign{\hfil$\scriptstyle##$\hfil\cr<\cr\sim\cr}}}
{\vcenter{\offinterlineskip\halign{\hfil$\scriptscriptstyle##$\hfil\cr<\cr\sim\cr}}}}}

\def\H{\mbox{\rm H}}

\def\Ma{\mbox{\rm Ma}}

\def\Pm{\mbox{\rm Pr}_{\rm M}}
\def\Rm{\mbox{\rm Re}_{\rm M}}

\def\Rey{\mbox{\rm Re}}

\def\EEM{{\cal E}_{\rm M}}

\def\EK{E_{\rm K}}
\def\EV{E_{\rm V}}
\def\EM{E_{\rm M}}

\def\kff{s}
\def\kffN{\mathcal{S}}
\def\qzeta{\zeta}
\def\tff{t_{\rm ff}}
\def\tph{t_{\rm ph}}

\def\cs{c_{\rm s}}

\def\EM{E_{\rm M}}

\def\Brms{B_{\rm rms}}
\def\calBrms{\mathcal{B}_{\rm rms}}

\def\urms{u_{\rm rms}}

\def\orms{\omega_{\rm rms}}

\def\half{{\textstyle{1\over2}}}

\def\onethird{{\textstyle{1\over3}}}

\newcommand{\W}{\,{\rm W}}
\newcommand{\V}{\,{\rm V}}

\newcommand{\T}{\,{\rm T}}
\newcommand{\G}{\,{\rm G}}

\newcommand{\K}{\,{\rm K}}
\newcommand{\g}{\,{\rm g}}
\newcommand{\s}{\,{\rm s}}

\newcommand{\m}{\,{\rm m}}

\newcommand{\J}{\,{\rm J}}

\newcommand{\A}{\,{\rm A}}

\def\apjl{ApJL}
\def\apjs{ApJS}

\def\aap{A\&A}

\begin{document}

\title{Magnetic field amplification during a turbulent collapse}

\author[0000-0002-7304-021X]{Axel Brandenburg}
\affiliation{Nordita, KTH Royal Institute of Technology and Stockholm University, Hannes Alfv\'ens v\"ag 12, SE-10691 Stockholm, Sweden}
\affiliation{The Oskar Klein Centre, Department of Astronomy, Stockholm University, AlbaNova, SE-10691 Stockholm, Sweden}
\affiliation{McWilliams Center for Cosmology \& Department of Physics, Carnegie Mellon University, Pittsburgh, PA 15213, USA}
\affiliation{School of Natural Sciences and Medicine, Ilia State University, 3-5 Cholokashvili Avenue, 0194 Tbilisi, Georgia}

\author[0000-0002-4324-0034]{Evangelia Ntormousi}
\affiliation{Scuola Normale Superiore, Piazza dei Cavalieri 7, 56126 Pisa, Italy}

\begin{abstract}
The question of whether a dynamo can be triggered by gravitational
collapse is of great interest, especially for the early Universe.
Here, we employ supercomoving coordinates to study the magnetic field
amplification from decaying turbulence during gravitational collapse.
We perform three-dimensional simulations and show that for large magnetic
Reynolds numbers, there can be exponential growth of the comoving magnetic
field with conformal time before the decay of turbulence impedes further
amplification.
The collapse dynamics only affect the nonlinear feedback from the
Lorentz force, which diminishes more rapidly for shorter collapse times,
allowing nearly kinematic continued growth.
We confirm that helical turbulence is more efficient in driving
dynamo action than nonhelical turbulence, but this difference decreases
for larger collapse times.
We also show that for nearly irrotational flows, dynamo amplification
is still possible, but it is always associated with a growth of
vorticity---even if it still remains very small.
In nonmagnetic runs, the growth of vorticity is associated with
viscosity and grows with the Mach number.
In the presence of magnetic fields, vorticity emerges from the curl of
the Lorentz force.
During a limited time interval, an exponential growth of the comoving magnetic
field with conformal time is interpreted as clear evidence of dynamo action.
\end{abstract}

\section{Introduction}

The hypothesis that dynamo action is ubiquitous in astrophysical plasmas
was introduced in the 1950s, but it faced skepticism due to various
antidynamo theorems \citep{Cow33, Hide+Palmer82}.
While initially the community focused on large-scale dynamos in the Sun
\citep{Par55, SKR66} and galaxies \citep{Par71, VR71}, the advance of
powerful computers brought significant attention to small-scale dynamos at
the scale of turbulence; see \cite{Meneguzzi+81} for the first
simulations and \cite{Kaz68} for the underlying theory, as well as
\cite{Kulsrud+Anderson92} for an independent and more detailed derivation.
By now, it is clear that three-dimensional turbulence always leads
to dynamo action when the plasma is sufficiently well conducting;
see \cite{BN23} for a recent review.
This behavior implies that part of the kinetic energy in the turbulence
is almost always converted into magnetic energy.

Collapse flows are particularly compelling for dynamo action.
Since gravitational collapse provides a strong source of kinetic energy,
it can enhance the magnetization of collapsing structures by sustaining
or introducing turbulence in the flow.
This mechanism is very relevant for galactic magnetism.
Recently, there have been claims of strong ($\sim\mu$G or stronger)
large-scale coherent galactic magnetic fields at redshifts up to 5.6
\citep{Geach2023, Chen2024}.
Assuming only tiny primordial seed magnetic fields, there might not be
enough time for a high-redshift galaxy to build strong enough magnetic
fields through mean-field dynamo action.
An early amplification of a tiny initial seed through
a small-scale dynamo \citep{Beck+94}, especially during the gravitational
collapse of the initial halo, could alleviate this problem.

Another relevant situation is star formation in the early Universe.
Primordial molecular clouds with initially negligible magnetic fields can become increasingly magnetized as they collapse,
an effect that is known to play a crucial role in the star formation process \citep{Pattle23}.

Despite its relevance to various astrophysical environments, gravitational
collapse dynamos have not yet been convincingly demonstrated.
The main reason is that characterizing dynamos in unsteady flows is inherently challenging.
For steady flows, we can always formulate an eigenvalue problem, provided the magnetic field is still weak and unaffected by the feedback from the Lorentz force, which affects the flow amplitude. It is even possible to prove that there is no eigenfunction with a nonvanishing eigenvalue when the magnetic diffusivity is strictly zero \citep{MP85}. Unsteady flows present a significant complication because, in that situation, the kinematic growth or decay of the magnetic field is no longer exponential.
The problem becomes approachable if the flow is statistically steady, i.e., the level of turbulence remains constant over time. In such cases, the energy spectrum grows at all wavenumbers at the same rate \citep{SB14}. This behavior is suggestive of the existence of an eigenfunction of the type discussed by \cite{Kaz68}. 
However, many astrophysical flows, such as gravitational collapse, are not even statistically steady.
Dynamo research in these cases is still in its infancy.

In a series of numerical simulations of isolated turbulent collapsing
molecular clouds, \citet{Sur+10, Sur+12} and \citet{Federrath+11b}
reported a significant amplification of the magnetic field.
However, in the absence of a proper criterion for dynamo action due to
the inherent difficulties described above, these works defined dynamo
action as any excess growth above the field $B\propto\rho^{2/3}$ expected
by gravitational collapse as the density $\rho$ increases.
Other works studying magnetic field growth in collapse flows
\citep[e.g.,][]{Schober2012,XL20} explicitly integrated the evolution of
the magnetic field through a turbulent dynamo.

A common problem faced in collapse simulations is to identify dynamo
action when other amplification mechanisms, such as tangling or
compression, are also active.
In this context, we proposed a criterion for dynamo action in unsteady
flows based on the work done against the Lorenz force \citep{BN22}.
Furthermore, by calculating the work against various forces, we emphasized
that the Jeans instability drives predominantly irrotational motions,
which are unlikely to account for any dynamo action observed in our simulation,
except for an early period before the collapse becomes more significant.

Kinetic helicity---a measure of the alignment between velocity and vorticity---is not necessary for dynamo action.
However, if present, it lowers the critical conductivity needed to overcome the effects of Joule dissipation \citep{Gilbert+88}.
Otherwise, resistive losses prematurely convert magnetic energy into
heat before the field can reach sufficient strength.

A collapsing flow can produce vorticity through viscosity
(especially in shocks), the baroclinic term, and magnetic
fields. However, which of these processes is active during collapse is currently unknown. 
To isolate the effects related to the collapse dynamics, \cite{Irshad+25} employed the supercomoving coordinates of \cite{Shandarin80},
where the conformal time $t$ is related to the physical time $\tph$ through $dt=d\tph/a^2$, and $a(t)$ is the scale factor;
see also \cite{MS98} for a detailed presentation of the supercomoving coordinates in magnetohydrodynamics.
\cite{Irshad+25} employed a supercomoving coordinate system that follows the self-gravitating collapse.
These coordinates enabled them to maintain sufficient numerical resolution
throughout the entire collapse, which is another common problem faced
in collapse simulations, including ours of 2022.

\cite{Irshad+25} found superexponential growth of the magnetic field as
a result of the increasing turnover rate and saturation field strengths
over the expectations from flux freezing. They applied a solenoidal forcing function with and without kinetic helicity.
The present work aims to study decaying turbulence
during gravitational collapse by employing supercomoving coordinates and
allowing not only for cases without initial kinetic helicity but also
cases with or without initial vorticity, i.e., acoustic turbulence.

\section{Our model}

\subsection{Supercomoving coordinates}

We employ supercomoving coordinates using the same definition of the
scale factor as \cite{Irshad+25}, i.e.,
\begin{equation}
a(t)=(1+\kff^2 t^2/4)^{-1},
\label{ascale}
\end{equation}
where $t$ is the conformal time, $\kff$ is a freefall parameter,
which is related to the freefall time $\tff=\pi/2\kff$.
The physical time $t_\mathrm{ph}$ is then given by
\begin{equation}
t_\mathrm{ph}(t)=\int_0^t a^2(t')\,\dd t',
\label{tph}
\end{equation}
which is defined in the range $0\leq t_\mathrm{ph}\leq \tff$.

The supercomoving coordinates stretch the finite time singularity at
$\tff$ to infinity while also limiting the comoving magnetic field
strength according to
\begin{equation}
B=a^2 B_\mathrm{ph},
\end{equation}
where $B_\mathrm{ph}$ is the physical magnetic field.

\subsection{Governing equations}

We solve the MHD equations with an isothermal equation of state, where
the pressure $p$ and density $\rho$ are related to each other through
$p=\rho\cs^2$ with $\cs=\const$ being the isothermal sound speed.
We apply an initial velocity field $\uu$, which leads to a turbulent
evolution.
We also apply an initial seed magnetic field $\BB$.
To ensure that $\BB$ remains solenoidal, we solve
for the magnetic vector potential $\AAA$ so that $\BB=\nab\times\AAA$.
The evolution equations for $\AAA$, $\uu$, and $\rho$ are given by
\begin{equation}
\frac{\partial\AAA}{\partial t}=\uu\times\BB+\eta\nabla^2\AAA,
\label{dAdt}
\end{equation}
\begin{equation}
\frac{\DD\uu}{\DD t}=-\cs^2\nab\ln\rho
+\rho^{-1}\left[a(t)\,\JJ\times\BB+\nab\cdot(2\nu\rho\SSSS)\right],
\label{DuDt}
\end{equation}
\begin{equation}
\frac{\DD\ln\rho}{\DD t}=-\nab\cdot\uu,
\label{DlrDt}
\end{equation}
where $\JJ=\nab\times\BB/\mu_0$ is the current density with $\mu_0$
being the vacuum permeability, $\JJ\times\BB$ is the Lorentz force,
$\SSSS$ the rate-of-strain tensor with the components ${\sf S}_{ij}=
\half(\partial_i u_j+\partial_j u_i)-\onethird\delta_{ij}\nab\cdot\uu$
and $\nu$ is the kinematic viscosity.

\subsection{Initial conditions and parameters}

We consider a cubic domain of size $L^3$ with periodic boundary conditions.
The lowest wavenumber in the domain is then $k_1\equiv2\pi/L$.
Owing to the use of periodic boundary conditions, the mass in the domain
is conserved, so the mean density is a constant, which defines our
reference density $\rho_0\equiv\meanrho$.
In the numerical simulations, we set $\cs=k_1=\rho_0=1$.

We construct our initial velocity in Fourier space (indicated by a tilde)
as $\tilde{\uu}(\kk)=\bm{{\mathsf{M}}}(\kk)\SSS(\kk)$. Here
\begin{equation}
S_j(\kk)=r(\kk,j)\,\frac{k_0^{-3/2} (k/k_0)}
{1+(k/k_0)^{17/6}},
\end{equation}
where $r(\kk,j)$ is a Gaussian-distributed random number with zero mean
and a variance of unity for each value of $\kk$ and each direction $j$,
$k_0$ is the peak wavenumber of the initial condition,
and $\bm{{\mathsf{M}}}$ is a matrix that consists of a superposition of
a vortical and an irrotational contribution \citep{BS25}:
\begin{equation}
\mathsf{M}_{ij}(\kk)=
(1-\qzeta)(\delta_{ij}-\hat{k}_i\hat{k}_j+\sigma\ii\hat{k}_k\epsilon_{ijk})
+\qzeta\hat{k}_i\hat{k}_j,
\label{Sfunc}
\end{equation}
where $0\leq \qzeta\leq1$ quantifies the irrotational fraction and
$0\leq\sigma\leq1$ the helicity fraction.
The extreme cases $\qzeta=0$ and $\qzeta=1$ correspond to vortical and irrotational
flows, respectively, while $\sigma=0$ and $\sigma=1$ correspond to
nonhelical and helical fields, respectively.
The shell-integrated kinetic energy spectrum, $\EK(k)$, which is
normalized such that $\int\EK(k)\,\dd k=\rho_0\bra{\uu^2/2}$, is
initially $\propto k^4$ for $k<k_0$ and $\propto k^{-5/3}$ for $k>k_0$.
The magnetic energy spectrum $\EM(k)$ is normalized such that
$\int\EM(k)\,\dd k=\bra{\BB^2/2\mu_0}$ and initially of the same
shape as $\EK(k)$.
We also compute the vortical energy spectrum $\EV(k)$, which is
normalized such that $\int k^2\EV(k)\,\dd k=\rho_0\bra{\oo^2/2}$,
where $\oo=\nab\times\uu$ is the vorticity.

It is often convenient to express our results not in code units,
where $\cs=k_1=\rho_0=1$, but in units of $u_0$ and $k_0$.
Here, $u_0\equiv\bra{\uu^2}^{1/2}$ is the initial rms velocity.
We also define a nondimensional magnetic field as
\begin{equation}
\mathcal{B}_i\equiv B_i/(\mu_0\rho_0 u_0^2)^{1/2},
\end{equation}
where $i=x,y,z$ refers to the three components, and $i=\mathrm{rms}$
or $i=\mathrm{ini}$ refer to the rms values of the magnetic field at
the actual or the initial time, respectively.
We also define the Mach and magnetic Reynolds numbers based on the initial
velocity, $\Ma_0=u_0/\cs$ and $\Rm=u_0/\eta k_0$, respectively.
The Mach number at the actual time is denoted by $\Ma$.
As a nondimensional measure of $\kff$, we define $\kffN=\kff/u_0 k_0$.
When $\kffN<1$ ($\kffN>1$), the collapse is slower (faster) than the
turnover rate of the turbulence.

In the following, we vary the input parameters $\kff$, $\qzeta$,
$k_0/k_1$, $\Ma_0$, $\Rm$, and $\mathcal{B}_\mathrm{ini}$.
In all cases presented below, the magnetic Prandtl number is unity,
i.e., $\nu/\eta=1$.

In the following, we display the conformal time in units of
the initial turnover time, $(u_0 k_0)^{-1}$, where $u_0$ is
the initial rms velocity.
As in \cite{BN22}, we monitor the vortical and irrotational
contributions to the turbulence, $\orms=\bra{\oo^2}^{1/2}$ and
$(\nab\cdot\uu)_\mathrm{rms}=\bbra{(\nab\cdot\uu)^2}^{1/2}$, in terms
of quantities that have the dimension of a wavenumber,
\begin{equation}
k_{\,\nab\cdot\uu}=(\nab\cdot\uu)_{\rm rms}/\urms,
\end{equation}
\begin{equation}
k_{\oo}=\orms/\urms.
\end{equation}
These two values are expected to scale with $k_0$, which is why we
usually present the ratios $k_{\,\nab\cdot\uu}/k_0$ and $k_{\oo}/k_0$.

\begin{table*}[htb]\caption{
Summary of the runs discussed in this paper.
Here we list the nondimensional parameter $\kffN$; the physical values
in code units are $\kff/\cs k_1=0.2$, 1, 5, 20, and 100
both for the helical and nonhelical runs, 1--7 and 8--14, respectively.
Column~7 gives $\Rm$ ($\Rey$) for magnetic (nonmagnetic) runs.
Dashes in columns~8--10 indicate the 8 nonmagnetic runs.
For magnetic runs, dashes in columns~9 and 10 indicate decay.
Run~39 corresponds to Run~B of \cite{BN22} and is discussed in
\Sec{HomogeneousCollapse}.
}\hspace{-20mm}\vspace{12pt}\centerline{\begin{tabular}{ccc ccc ccc ccc c}
Run& $\kffN$ & $\sigma$ & $\qzeta$ & $k_0/k_1$ & $\Ma_0$ & $\Rm$ ($\Rey$) & $\mathcal{B}_\mathrm{ini}$ & $\Delta\ln\mathcal{B}$ &
$\lambda/u_0 k_0$ & $\Delta\ln(k_\omega/k_0)$ & $(k_\omega/k_0)_{\max}$ & resol.\ \\
\hline
 1 &  0.1 &  1 &  0   & 10 & 0.18 & 1840  &$2.3\times10^{-8}$ &  8.33 &  0.52 &  0.39 &  7.09 &$ 512^3$ \\
 2 &  0.1 &  1 &  0   & 10 & 0.18 & 1840  &$2.3\times10^{-5}$ &  6.62 &  0.52 &  0.39 &  7.09 &$ 512^3$ \\
 3 &  0.1 &  1 &  0   & 10 & 0.18 & 1840  &$2.3\times10^{-2}$ &  1.88 &  1.00 &  0.31 &  6.46 &$ 512^3$ \\
 4 &  0.6 &  1 &  0   & 10 & 0.18 & 1840  &$2.3\times10^{-2}$ &  2.21 &  1.03 &  0.22 &  5.93 &$ 512^3$ \\
 5 &  2.8 &  1 &  0   & 10 & 0.18 & 1840  &$2.3\times10^{-2}$ &  3.56 &  1.03 &  0.30 &  6.43 &$ 512^3$ \\
 6 &  11  &  1 &  0   & 10 & 0.18 & 1840  &$2.3\times10^{-2}$ &  4.77 &  1.03 &  0.36 &  6.82 &$ 512^3$ \\
 7 &  56  &  1 &  0   & 10 & 0.18 & 1840  &$2.3\times10^{-2}$ &  5.96 &  1.03 &  0.39 &  7.04 &$ 512^3$ \\
\hline
 8 &  0.2 &  0 &  0   & 10 & 0.13 & 1300  &$3.3\times10^{-8}$ &  4.27 &  0.37 &  0.33 &  6.97 &$ 512^3$ \\
 9 &  0.2 &  0 &  0   & 10 & 0.13 & 1300  &$3.3\times10^{-5}$ &  4.22 &  0.37 &  0.33 &  6.97 &$ 512^3$ \\
10 &  0.2 &  0 &  0   & 10 & 0.13 & 1300  &$3.3\times10^{-2}$ &  1.49 &  0.97 &  0.14 &  5.70 &$ 512^3$ \\
11 &  0.8 &  0 &  0   & 10 & 0.13 & 1300  &$3.3\times10^{-2}$ &  1.92 &  0.97 &  0.17 &  5.91 &$ 512^3$ \\
12 &  3.8 &  0 &  0   & 10 & 0.13 & 1300  &$3.3\times10^{-2}$ &  3.03 &  0.98 &  0.29 &  6.66 &$ 512^3$ \\
13 &  15  &  0 &  0   & 10 & 0.13 & 1300  &$3.3\times10^{-2}$ &  3.75 &  0.98 &  0.33 &  6.92 &$ 512^3$ \\
14 &  77  &  0 &  0   & 10 & 0.13 & 1300  &$3.3\times10^{-2}$ &  4.12 &  0.98 &  0.33 &  6.97 &$ 512^3$ \\
\hline
15 &  0.2 &  0 & 0.10 & 10 & 0.12 & 1170  &$3.6\times10^{-2}$ &  1.41 &  0.34 &  0.11 &  5.50 &$ 512^3$ \\
16 &  0.2 &  0 & 0.50 & 10 & 0.08 &  800  &$5.4\times10^{-2}$ &  1.04 &  0.25 &  0.00 &  4.00 &$ 512^3$ \\
17 &  0.2 &  0 & 0.90 & 10 & 0.08 &  840  &$5.1\times10^{-2}$ &  0.31 &  0.04 &  0.25 &  0.94 &$ 512^3$ \\
18 &  0.2 &  0 & 0.95 & 10 & 0.09 &  880  &$4.9\times10^{-2}$ &  0.05 & 0.003 &  0.28 &  0.47 &$ 512^3$ \\
19 &  0.2 &  0 & 0.96 & 10 & 0.09 &  880  &$4.8\times10^{-2}$ &  0.02 & 0.001 &  0.26 &  0.38 &$ 512^3$ \\
20 &  0.2 &  0 & 0.97 & 10 & 0.09 &  890  &$4.8\times10^{-2}$ &  ---  &  ---  &  0.21 &  0.29 &$ 512^3$ \\
21 &  0.2 &  0 & 0.98 & 10 & 0.09 &  900  &$4.7\times10^{-2}$ &  ---  &  ---  &  0.13 &  0.20 &$ 512^3$ \\
22 &  0.2 &  0 & 0.99 & 10 & 0.09 &  910  &$4.7\times10^{-2}$ &  ---  &  ---  &  0.20 &  0.16 &$ 512^3$ \\
23 &  0.2 &  0 &  1   & 10 & 0.09 &  920  &$4.6\times10^{-2}$ &  ---  &  ---  &  0.30 &  0.14 &$ 512^3$ \\
\hline
24 &  0.1 &  0 &  1   & 20 & 0.09 &  920  &     ---           &  ---  &  ---  &  0.01 &  0.07 &$1024^3$ \\
25 &  0.2 &  0 &  1   & 10 & 0.09 &  930  &     ---           &  ---  &  ---  &  0.03 &  0.05 &$ 1024^3$ \\
26 &  0.4 &  0 &  1   &  5 & 0.09 &  940  &     ---           &  ---  &  ---  &  0.38 &  0.04 &$ 1024^3$ \\
27 &  1.0 &  0 &  1   &  2 & 0.10 &  950  &     ---           &  ---  &  ---  &  1.27 &  0.03 &$ 1024^3$ \\
\hline
28 &  0.5 &  0 & 0.95 & 10 & 0.04 &  220  &     ---           &  ---  &  ---  &  0.09 &  0.23 &$ 512^3$ \\
29 &  0.1 &  0 & 0.95 & 10 & 0.18 &  890  &     ---           &  ---  &  ---  &  0.31 &  0.71 &$ 1024^3$ \\
30 &  0.1 &  0 & 0.95 & 10 & 0.27 & 1330  &     ---           &  ---  &  ---  &  0.43 &  1.00 &$ 1024^3$ \\
31 &  0.1 &  0 & 0.95 & 10 & 0.36 & 1780  &     ---           &  ---  &  ---  &  0.51 &  1.31 &$ 1024^3$ \\
\hline
32 &  0.2 &  0 & 0.96 & 10 & 0.09 &  900  &$4.9\times10^{-2}$ &  0.02 & 0.001 &  0.17 &  0.38 &$ 1024^3$ \\
33 &  0.2 &  0 & 0.96 & 10 & 0.09 & 1800  &$4.9\times10^{-2}$ &  0.12 & 0.004 &  0.28 &  0.53 &$ 1024^3$ \\
34 &  0.2 &  0 & 0.96 & 10 & 0.09 & 4500  &$4.9\times10^{-2}$ &  0.51 & 0.008 &  0.53 &  0.79 &$ 1024^3$ \\
\hline
35 &  0.2 &  0 &  1   & 10 & 0.09 & 1870  &$9.4\times10^{-3}$ &  ---  &  ---  &  0.03 &  0.07 &$ 1024^3$ \\
36 &  0.2 &  0 &  1   & 10 & 0.09 & 1870  &$2.4\times10^{-2}$ &  ---  &  ---  &  0.17 &  0.09 &$ 1024^3$ \\
37 &  0.2 &  0 &  1   & 10 & 0.09 & 1870  &$4.7\times10^{-2}$ &  ---  &  ---  &  0.34 &  0.21 &$ 1024^3$ \\
38 &  0.2 &  0 &  1   & 10 & 0.09 & 1870  &$9.4\times10^{-2}$ &  ---  &  ---  &  0.25 &  0.48 &$ 1024^3$ \\
\hline
39 &  0.4 &  1 &  0   & 10 & 0.19 &  190  &$2.3\times10^{-17}$&  8.32 &  0.42 &  0.01 &  4.29 &$ 2048^3$ \\
\label{Tsummary}\end{tabular}}\end{table*}

We use for all simulations the \textsc{Pencil Code} \citep{PC}.
Except for Run~39, where the resolution is $2048^3$ mesh points, it is
either $512^3$ or $1024^3$, as indicated in
\Tab{Tsummary}, where we summarize all runs discussed in this paper.
As discussed later in \Sec{CriticalVorticity}, $k_\omega/k_0$ starts off
with a small, but finite value, decreases rapidly at first, and may later display
a continuous growth until a maximum $(k_\omega/k_0)_{\max}$ is reached.
When a maximum is reached, we denote the total growth in $e$-folds from minimum to maximum by
$\Delta\ln(k_\omega/k_0)$, which is analogous to the growth in $e$-folds
of the magnetic field, which we denote by $\Delta\ln\mathcal{B}$.

While higher resolution leads to more accurate results, the lower
resolution computations produce qualitatively similar ones;
compare, for example, Runs~19 and 32, which have the same parameters.
Both runs have almost the same vorticity and magnetic field evolution, but
the lower resolution run has a slightly deeper minimum of $k_\omega/k_0$,
which results in a larger value of $\Delta\ln(k_\omega/k_0)$.

\section{Results}

\subsection{Growth versus physical and conformal time}
\label{GrowthConformal}

\begin{figure}[t!]\begin{center}
\includegraphics[width=\columnwidth]{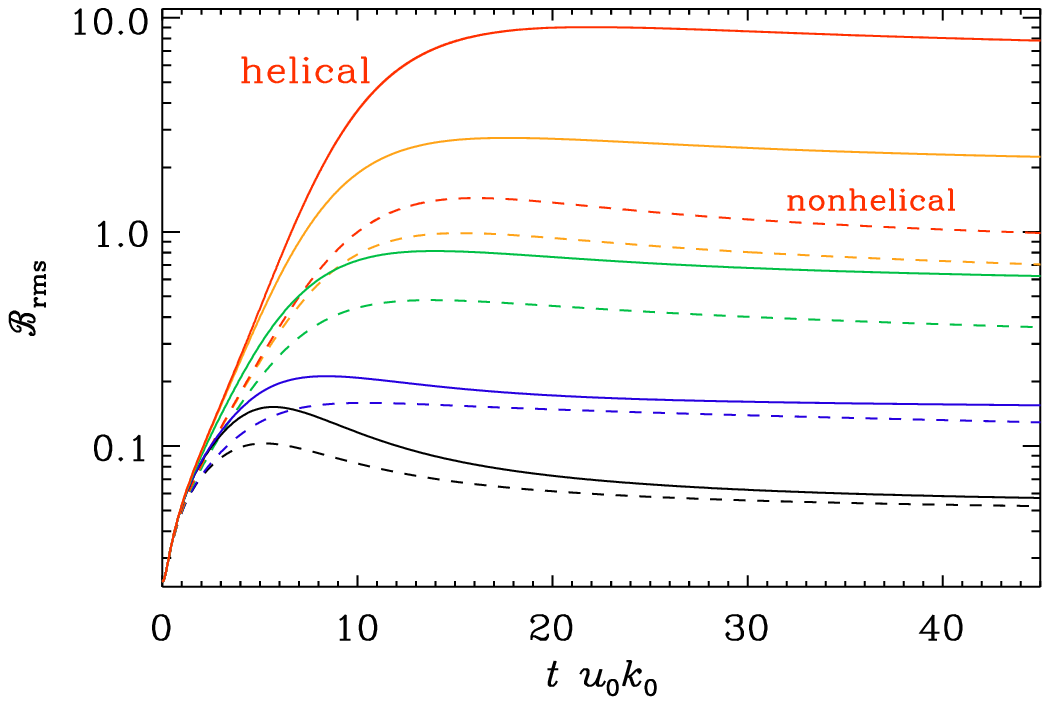}
\end{center}\caption{
Evolution of the rms magnetic field in comoving coordinates for
$\kff/\cs k_1=0.2$ (black lines), 
1 (blue lines),  
5 (green lines),  
20 (orange lines), and 
100 (red lines).  
Solid (dashed) lines refer to cases with (without)
initial kinetic helicity and have values of $u_0$ that are slightly
larger (smaller), so $\kffN$ is in the range 0.1--56 (0.2--77); see \Tab{Tsummary}.
Runs~3--7 and Runs~10--14. 
}\label{pcomp_hel_nhel}\end{figure}

We have performed runs with different values of $\kffN$ using either
helical ($\sigma=1$) or nonhelical ($\sigma=0$) turbulence, sometimes
without irrotational contributions ($\qzeta=0$).
\Fig{pcomp_hel_nhel} shows that the larger the value of $\kffN$, the
larger the final magnetic field strength.
This is because the effective Lorentz force in \Eq{DuDt}, $a\JJ\times\BB$,
diminishes more rapidly with time when $\kffN$ is larger, allowing the
magnetic field to continue growing further.
In supercomoving coordinates, the initial growth rate of the magnetic field
is not affected by the value of $\kffN$.
However, the growth rate is larger with than without kinetic helicity.
On the other hand, at later times, when the magnetic field decays, the
values are similar regardless of the presence of kinetic helicity.

\begin{figure}[t!]\begin{center}
\includegraphics[width=\columnwidth]{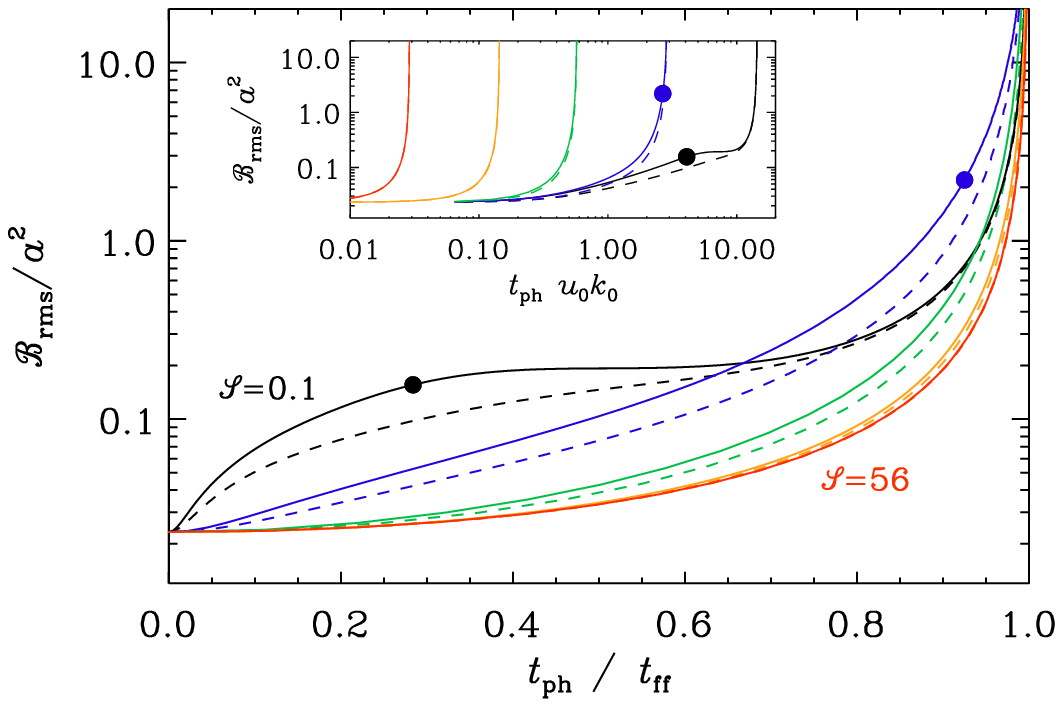}
\end{center}\caption{
Same as \Fig{pcomp_hel_nhel}, but in physical units.
Time is here normalized by the freefall time.
The black and blue dots on the black and blue curves denote the time until
which the growth in \Fig{pcomp_hel_nhel} was still approximately exponential.
The inset shows the same, but now time is normalized by the initial
turnover time.
Runs~3--7 and Runs~10--14. 
}\label{pcomp_hel_nhel_phys}\end{figure}

In physical time, the magnetic field shows a steep increase
just toward the end of the collapse; see \Fig{pcomp_hel_nhel_phys}.
Interestingly, the runs with large values of $\kffN$, which produce
the strongest comoving magnetic fields, now yield the weakest physical fields
when comparing the runs at the same fractional collapse time.
This is because for the runs with large values of $\kffN$,
the freefall time is short, so the fractional times are larger,
which effectively interchanges the order of the curves.
This is demonstrated in the inset of \Fig{pcomp_hel_nhel_phys},
where we show the same data, but now with time in units of
the initial turnover time.

In \Fig{pcomp_hel_nhel_phys}, we have also indicated the times where the
initial exponential growth of the comoving magnetic field with conformal
time terminates.
For $\kffN=0.1$ and 0.6, $\calBrms/a^2$ has hardly increased by an order
of magnitude.
In particular, the growth of $\calBrms/a^2$ versus physical time is not
superexponential, as found by \cite{Irshad+25}.
The reason for our subexponential growth for $\kffN\ll1$ is that the rms
velocity decreases significantly due to turbulent diffusion, leading to
a smaller growth rate, which then counters the effect of collapse.
Only for larger values of $\kffN$ is the growth superexponential in
physical coordinates, and exponential in comoving coordinates.
For $\kffN\geq2.8$, the times when exponential growth in
comoving coordinates terminates are outside the plot range of
\Fig{pcomp_hel_nhel_phys}.

\begin{figure}[t!]\begin{center}
\includegraphics[width=\columnwidth]{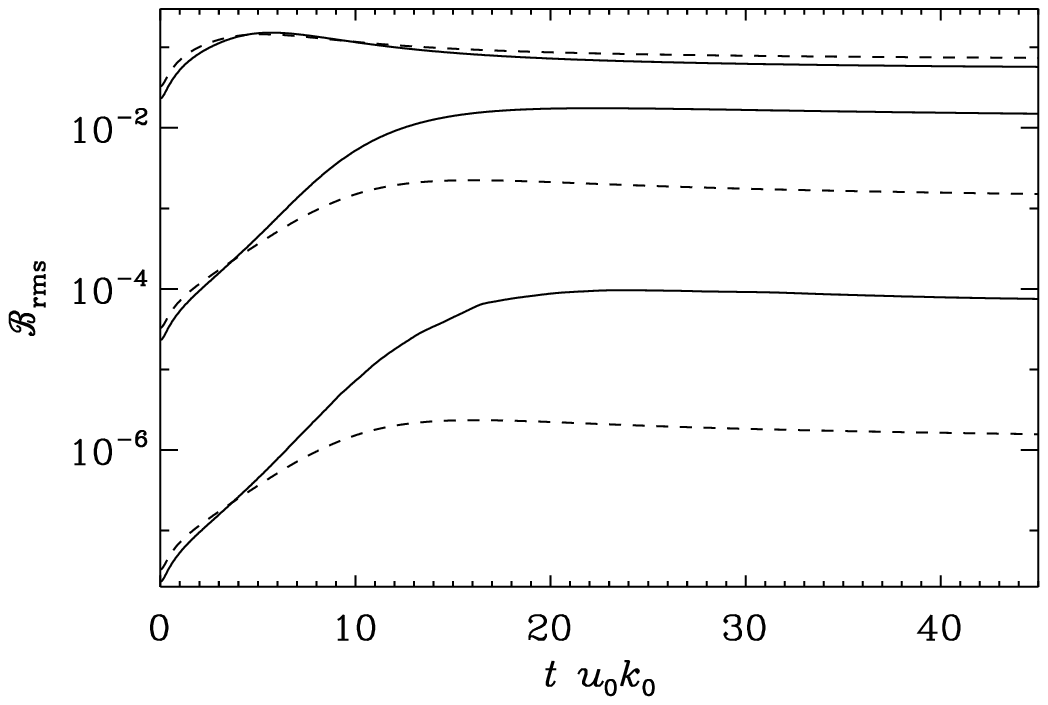}
\end{center}\caption{
Same as \Fig{pcomp_hel_nhel}, but for 3 different initial field strengths.
Runs~1--3 and Runs~8--10. 
}\label{pcomp_hel_nhel_all}\end{figure}

Given that the only effect of the collapse is on the Lorentz
force, it is clear that the kinematic phase is completely
independent of the collapse.
In the runs with a smaller initial field, the kinematic growth phases
can last longer before the turbulence has decayed too much, while for
a stronger initial field, nonlinear effects terminate the exponential
growth phase earlier.
This is shown quantitatively in \Fig{pcomp_hel_nhel_all},
where we see the magnetic field growth for different initial
field strengths.
For weak initial fields, the comoving magnetic field grows
by more than three orders of magnitude.
It could grow more strongly if the magnetic Reynolds number
were larger.
The growth is only limited by the competition between magnetic field
amplification by the flow and the simultaneous decay of the flow.
Similar results were already reported in \cite{Bra+19}, but without
collapse dynamics ($a=1$).

\subsection{Effect of the Lorentz force}

\begin{figure}[t!]\begin{center}
\includegraphics[width=\columnwidth]{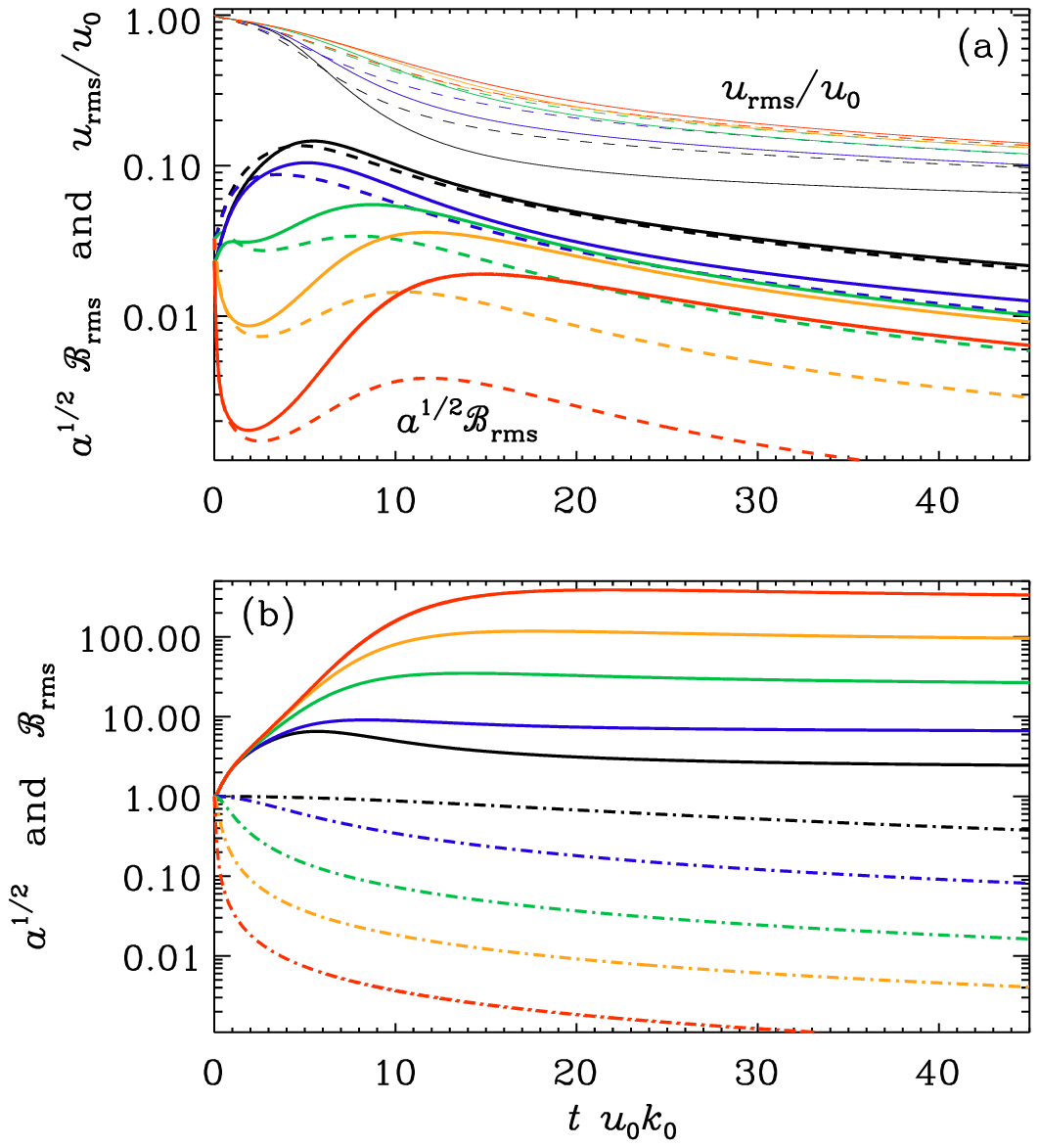}
\end{center}\caption{
(a) Similar to \Fig{pcomp_hel_nhel}, but now $a^{1/2}\calBrms$ (thicker
lines) and the instantaneous rms velocity (thinner lines) are plotted.
The order of the colors is the same as before, with black being for
$\kff/\cs k_1=0.2$ and red for $\kff/\cs k_1=100$ and solid (dashed) lines refer to
helical (nonhelical) initial flows,
for which $\kffN$ varies in the range 0.1--56 (0.2--77).
(b) Evolution separately for $a^{1/2}$ (dashed--dotted lines) and
$\calBrms$ (solid lines), again with the same colors as before.
Runs~3--7 and Runs~10--14. 
}\label{pcomp_hel_nhel_withu}\end{figure}

As we have seen from \Fig{pcomp_hel_nhel_all}, when the initial magnetic
field strength is large, the early exponential growth diminishes more
rapidly.
This is the result of the effective Lorentz force in \Eq{DuDt} becoming
comparable with the inertial term, which implies \citep{Irshad+25}
\begin{equation}
a^{1/2}\Brms\la\urms\sqrt{\mu_0\rho_0}.
\label{Sat}
\end{equation}
This is demonstrated in \Fig{pcomp_hel_nhel_withu}(a), where we compare
the evolution of $a^{1/2}\calBrms$ with that of $\urms/u_0$ for the same
runs as those of \Figs{pcomp_hel_nhel}{pcomp_hel_nhel_phys}.

We see that \Eq{Sat} is well obeyed for all runs.
The largest values of $a^{1/2}\calBrms$ are obtained for the runs
with small values of $\kffN$.
The effect of kinetic helicity here is surprisingly weak, and the values
of $a^{1/2}\calBrms$ are only slightly smaller for the nonhelical runs than
for the helical ones.
For larger values of $\kffN$, on the other hand, the differences between
helical and nonhelical runs are much larger and we see that the decay
of $a^{1/2}$ is well overcompensated by the growth of $\calBrms$ so that
the product $a^{1/2}\calBrms$ still shows a strong increase later in the
evolution; see \Fig{pcomp_hel_nhel_withu}(b), where we plot separately
the evolutions of $a^{1/2}$ and $\calBrms$.

We also see that for large values of $\kffN$ (short freefall times),
$a^{1/2}\calBrms$ decays at early times and only shows growth after that.
This is opposite to the case of small values of $\kffN$ and simply
because at early times, $a^{1/2}$ decays faster than the exponential
growth of $\calBrms$.
Only somewhat later, for $2\la t u_0 k_0\la10$, exponential growth
prevails.

\subsection{Critical vorticity}
\label{CriticalVorticity}

Numerical simulations have demonstrated in the past that vorticity
is an important ingredient of dynamos \citep{HBM04, Federrath+11}.
\cite{AchikanathChirakkara+21} did report dynamo action for purely
irrotational driving, but this could perhaps still be explained
by some residual vorticity in their simulations.

\begin{figure}[t!]\begin{center}
\includegraphics[width=\columnwidth]{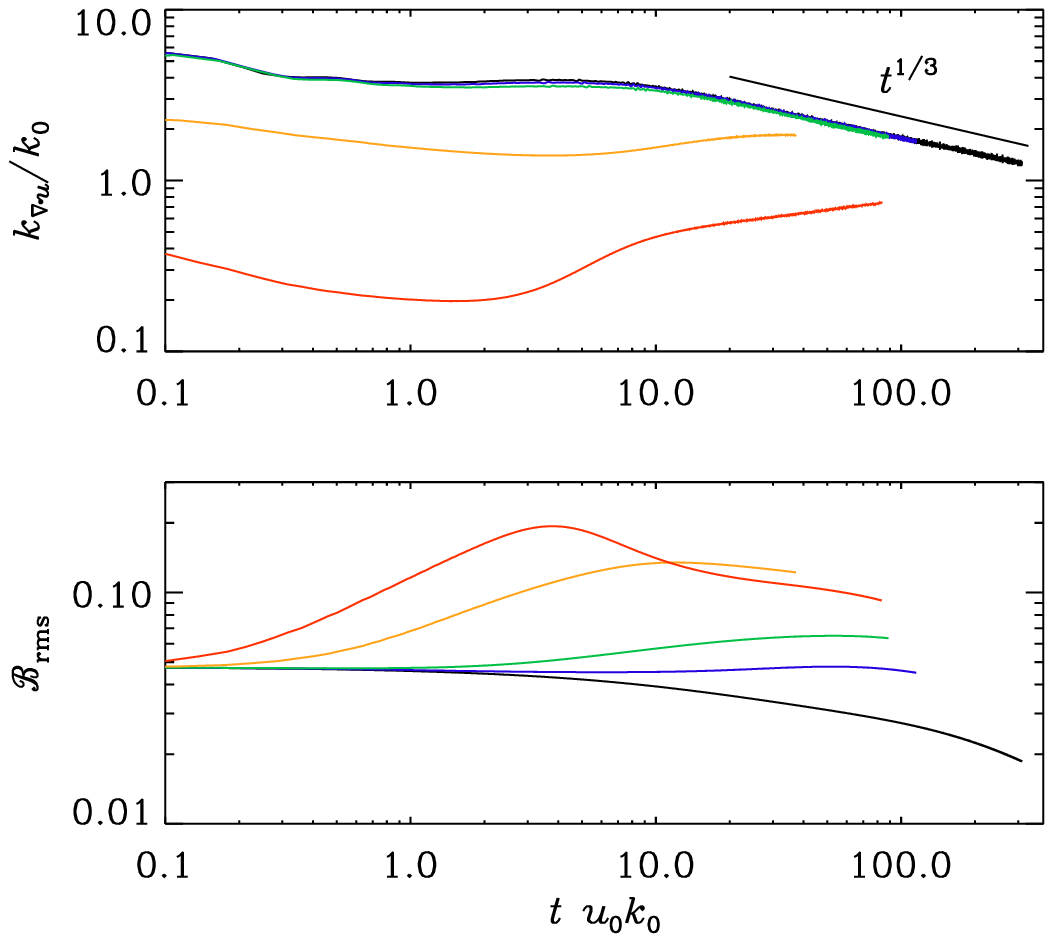}
\end{center}\caption{
Evolution of $k_{\nabla\cdot\uu}/k_0$ (upper panel) and $\calBrms$ (lower panel)
for $\qzeta=0.1$ (red), $0.5$ (orange), $0.9$ (green),
$0.95$ (blue), and $1$ (black).
Runs~15--18 and Run~23. 
}\label{pcomp_qirro}\end{figure}

The apparent necessity of vorticity may be a limitation of current
simulations, whose maximum magnetic
Reynolds number may still not be large enough, because theoretically,
small-scale dynamo action should also be possible for irrotational
turbulence \citep{Kazantsev+85, Afonso+19}.
Clarifying this question for collapse simulations with the effective gain
in resolution due to the use of supercomoving coordinates is crucial.
We can study this here in more detail by varying the value of $\qzeta$.
In \Fig{pcomp_qirro} we plot the evolution of $k_{\,\nab\cdot\uu}/k_0$
and $\calBrms$ for runs with $\Rm=900$ and several values of $\qzeta$.
It is only when $\qzeta$ is very close to unity that dynamo action ceases.
This suggests that very small amounts of vorticity can suffice for
successful dynamo action.
The intervals displaying a steady increase of $k_{\,\nab\cdot\uu}/k_0$, which were also seen in
the work of \cite{BN22}, are just a consequence of the more rapid decay
of $\urms$ compared to $(\nab\cdot\uu)_{\rm rms}$.
At early times, $\urms$ is approximately constant while
$(\nab\cdot\uu)_{\rm rms}$ shows an approximate power law decrease.
This explains the initial decrease of $k_{\,\nab\cdot\uu}/k_0$.

\begin{figure}[t!]\begin{center}
\includegraphics[width=\columnwidth]{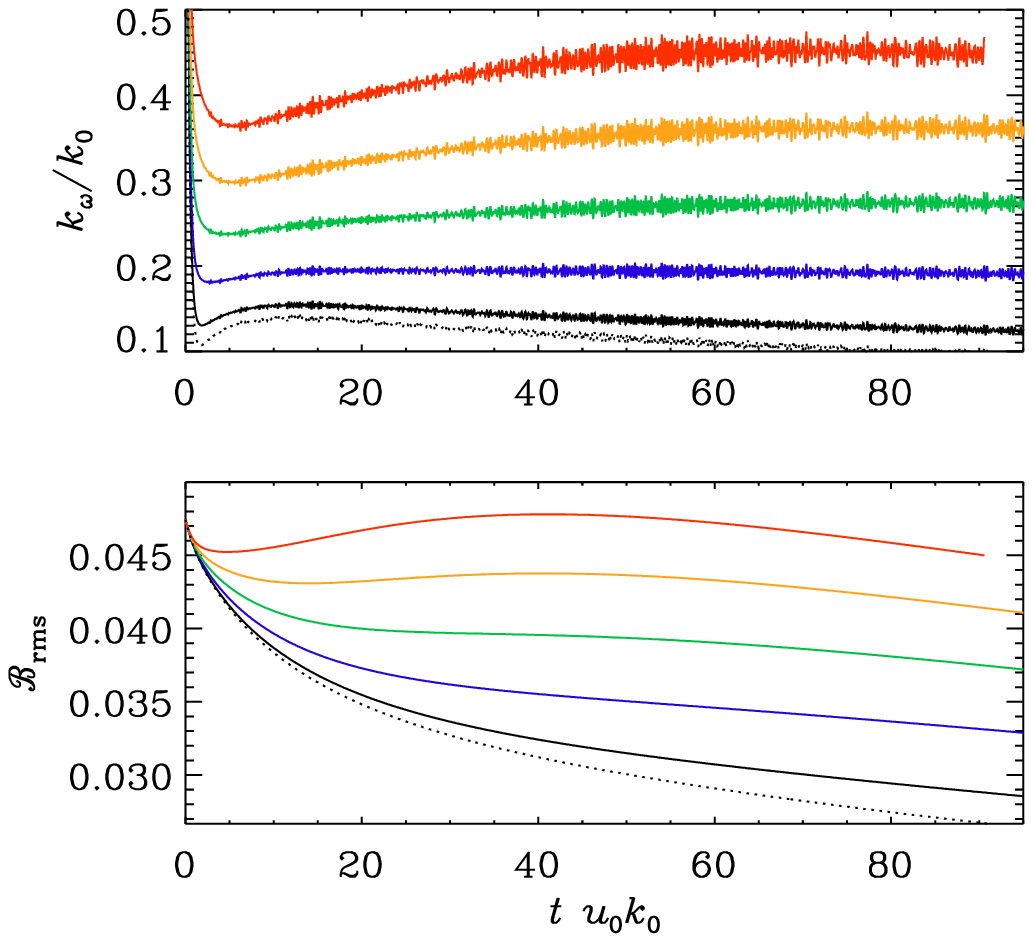}
\end{center}\caption{
$k_\omega/k_0$ (upper panel) and $\calBrms$ (lower panel)
for $1$ (dotted black), $0.99$ (solid black), $0.98$ (blue), 
$0.97$ (green), $0.96$ (orange), and $\qzeta=0.95$ (red).
Runs~18--23. 
}\label{pcomp_qirro2}\end{figure}

\begin{figure}[t!]\begin{center}
\includegraphics[width=\columnwidth]{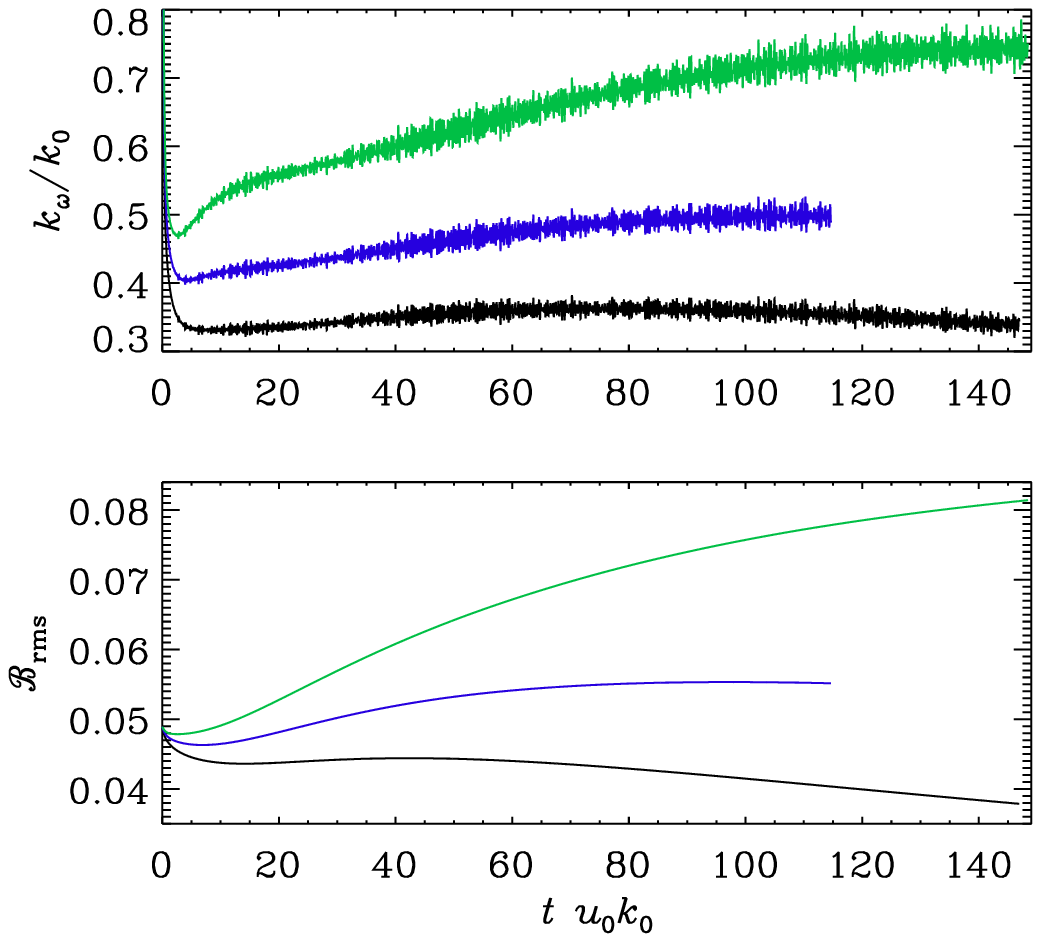}
\end{center}\caption{
$k_\omega/k_0$ (upper panel) and $\calBrms$ (lower panel)
for $\Rm=900$ (black), $1800$ (blue), and $4400$ (green).
The frequency of the oscillations is $\omega\approx15$.
The resolution is in all cases $1024^3$ mesh points.
Runs~32--34. 
}\label{pcomp_qirro3}\end{figure}

In \Fig{pcomp_qirro2} we focus on several more values of $\qzeta$ close to unity and
find that for $\Rm=880$, the critical value is around 0.96.
For larger values of $\qzeta$, there is no growth;
see Runs~20--23 and Runs~35--38.
However, the critical value of $1-\qzeta$ decreases with increasing
magnetic Reynolds number.
For larger values of $\Rm$, smaller amounts of vorticity suffice for
dynamo action.
This is shown in \Fig{pcomp_qirro3}, where we compare runs for
$\qzeta=0.96$ with different values of $\Rm=900$, 1800, and 4500, using
$1024^3$ mesh points.
This value of $\qzeta$ led to a vorticity that was the marginal value for
obtaining growing magnetic fields for $\Rm=900$.
We see that, as we increase $\Rm$, the episode of growth becomes longer
and the maximum magnetic field larger.

To assess the level of vorticity, 
it is of interest to define a Reynolds number based on the vorticity as
\citep{HBM04, Elias-Lopez+23, Elias-Lopez+24}
\begin{equation}
\Rey_\omega=\orms/\nu k_0^2,
\end{equation}
and compute the critical value above which dynamo action occurs.
Looking at \Tab{Tsummary}, we see that the threshold of $\qzeta$
between 0.96 and 0.97 corresponds to $k_\omega/k_0=0.38$ and 0.29,
respectively. With $\Rm\approx900$, the critical value is
$\Pm\,\Rey_\omega=(k_\omega/k_0)\,\Rm\approx300$.
This value is rather large, but it is unclear whether the dynamo onset
is indeed determined predominantly by $\Rey_\omega$.
If dynamos do indeed work for purely acoustic turbulence ($\qzeta=1$),
as found by \cite{AchikanathChirakkara+21}, the dynamo onset could
not depend on $\Rey_\omega$ alone.
Thus, future work should establish to what extent our critical value of
$\Pm\,\Rey_\omega$ of 300 is universal.

\subsection{Effect of scale separation}

\begin{figure}[t!]\begin{center}
\includegraphics[width=\columnwidth]{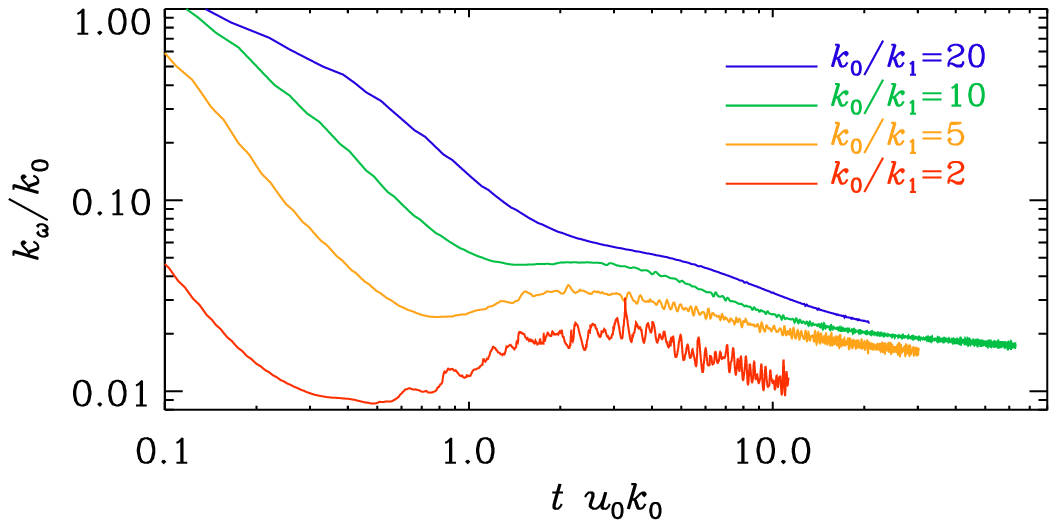}
\end{center}\caption{
$k_\omega/k_0$ for hydrodynamic runs with $\qzeta=1$,
$\Rm=900$, and different values of $k_0$.
Runs~24--27. 
}\label{pcomp_qirro2_hydro_k}\end{figure}

We have seen from \Fig{pcomp_qirro2} that for very small values of
$1-\qzeta$, the expected approach of $k_\omega$ to zero slows down
in the sense that the values are almost the same for $\qzeta=1$ and
$\qzeta=0.99$, and that for $\qzeta=0.98$ is further away.
It is conceivable that the finite value of $k_\omega$ for $\qzeta=1$
is caused by nonrepresentative averages resulting from a small number
of turbulent eddies, i.e., from small scale separation, which is the
ratio between $k_0$ and the lowest wavenumber of the domain.
To check whether this is the case, we present in \Fig{pcomp_qirro2_hydro_k}
runs with different values of $k_0$.
As expected, we see that $k_\omega$ scales with $k_0$, so the ratio
$k_\omega/k_0$ varies only little and lies in the range $0.01\leq
k_\omega/k_0\leq0.02$ after about 10--30 turnover times.
This suggests that this value of $k_\omega/k_0$ is not affected by the
finite scale separation.
When we decrease the scale separation ratio to $k_0/k_1=2$, the
run shows vigorous fluctuations.
They may indicate that the numerical resolution becomes insufficient
in the collapsing regions.
The above simulations have demonstrated once again that without the gain
of effective resolution due to the use of supercomoving coordinates,
earlier collapse simulations may have been severely underresolved.

\subsection{Growth of vorticity}

In \Fig{pcomp_qirro2}, we have seen that for $\qzeta=0.95$, there can be
growth of $k_\omega$ by a certain amount.
It is possible that this is caused either by magnetic driving
\citep{Kahn+12} or by what is known as magnetically assisted vorticity
production \citep{BS25}.
To clarify this, it is useful to compare with the purely hydrodynamic case;
see \Tab{Tsummary}.

For an isothermal gas, there is no baroclinic term, which would be the
main agent for producing vorticity in nonisothermal flows.
There is also no rotation or shear, both of which could lead to vorticity
generation \citep{DSB11, Elias-Lopez+23, Elias-Lopez+24}. There remain only
three possibilities for driving or amplifying vorticity: (i) through
viscosity via gradients of the velocity divergence being inclined against
density gradients, (ii) through magnetic driving or magnetically assisted
vorticity production \citep{BS25}, and (iii) through nonlinearity.

The growth of vorticity through nonlinearity may be motivated by the
formal analogy with the induction equation when the magnetic field is
replaced by the vorticity $\oo$, i.e.,
\begin{equation}
\frac{\partial\oo}{\partial t}=\nab\times(\uu\times\oo)
+\dot{\oo}_\mathrm{visc}+\dot{\oo}_\mathrm{mag},
\label{vort}
\end{equation}
where $\dot{\oo}_\mathrm{visc}=\nu(\nabla^2\oo+\nab\times\GG)$ is the curl
of the viscous acceleration with $G_i=2\mathsf{S}_{ij}\nabla_j\ln\rho$
being a vector characterizing the driving of vorticity
even if it was vanishing initially \citep{MB06,BS25}, and
$\dot{\oo}_\mathrm{mag}=a(t)\nab\times(\JJ\times\BB/\rho)$ is the vorticity
driving from the curl of the Lorentz force,
where we have included the $a(t)$ term resulting from the use of
supercomoving coordinates.

The analogy between induction and vorticity equations is obviously
imperfect, because the velocity is directly related to the vorticity.
This analogy has been invoked by \cite{Batchelor50} to explain dynamo
action, but here we rather use it to motivate the question whether
vorticity can be amplified.

\begin{figure}[t!]\begin{center}
\includegraphics[width=\columnwidth]{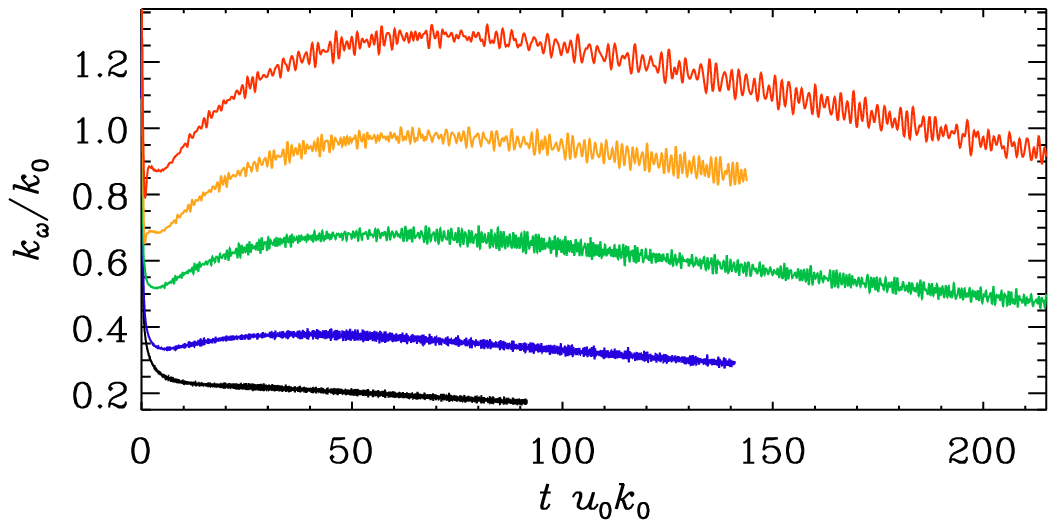}
\end{center}\caption{
Evolution of $k_\omega/k_0$ for different Mach numbers.
Runs~28--31. 
}\label{pcomp_qirro2_hydro_mach}\end{figure}

\begin{figure}[t!]\begin{center}
\includegraphics[width=\columnwidth]{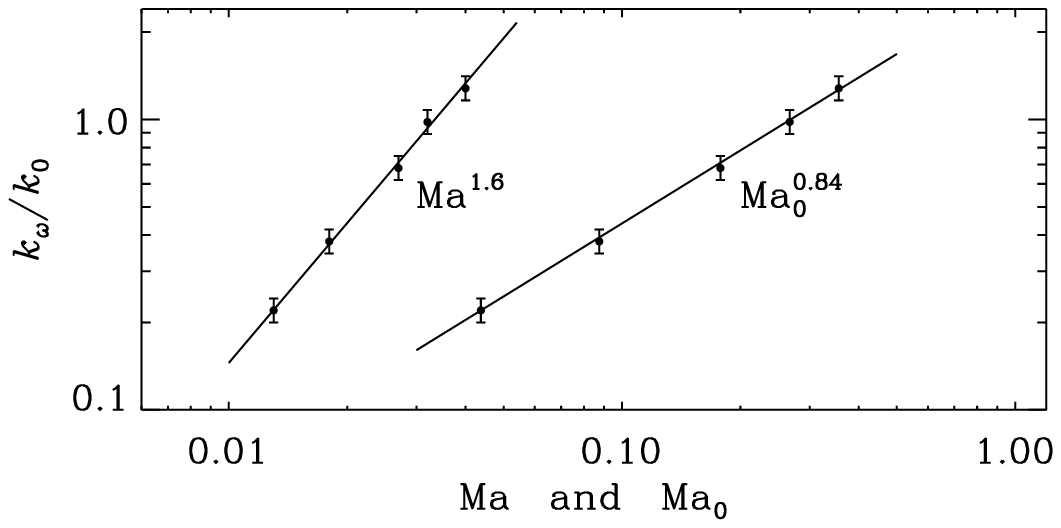}
\end{center}\caption{
Scaling of $(k_\omega/k_0)_{\max}$ with the actual and initial
Mach numbers, $\Ma$ and $\Ma_0$, respectively.
The slopes are 1.6 and 0.84, respectively.
Runs~28--31. 
}\label{pko_vs_mach}\end{figure}

To distinguish between the various possibilities, we must vary the
viscosity, the Mach number, and the initial magnetic field strength.
One important clue is given by the fact that the occurrence of vorticity
depends on the Mach number of the turbulence.
This is demonstrated in \Fig{pcomp_qirro2_hydro_mach}, where we plot
the evolution of $k_\omega/k_0$ for different Mach numbers.
\Fig{pko_vs_mach} shows that $(k_\omega/k_0)_{\max}$ scales with the actual 
Mach number $\Ma$ at the time when $(k_\omega/k_0)_{\max}$ is reached
and the initial Mach number $\Ma_0$, respectively.
The slopes for both scalings are different, and somewhat shallower than
the nearly quadratic scaling found by \cite{Federrath+11} for the forced case.

\begin{figure}[t!]\begin{center}
\includegraphics[width=\columnwidth]{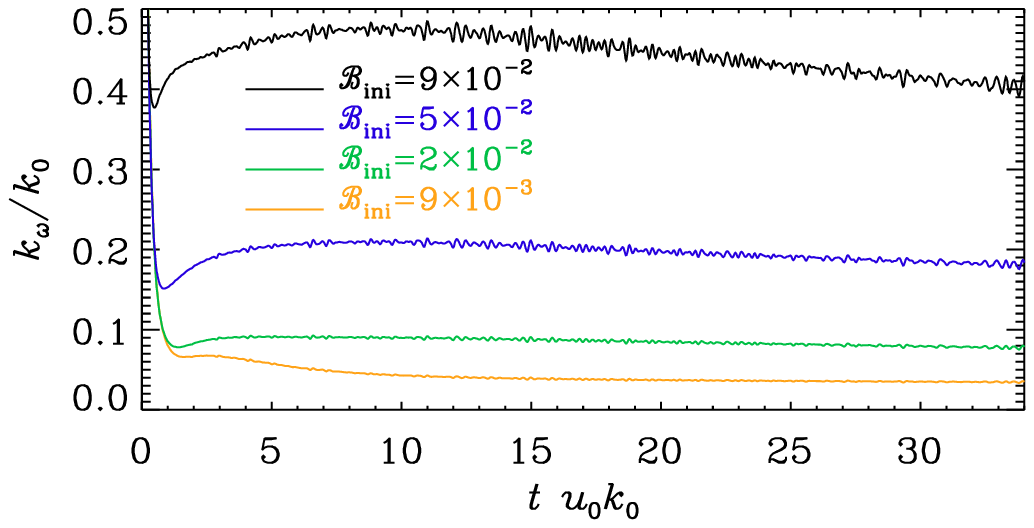}
\end{center}\caption{
$k_\omega/k_0$ for hydromagnetic runs with $\qzeta=1$,
$\Rm=1900$, and different magnetic field strengths.
Runs~35--38. 
}\label{pcomp_qirro2_hydro_B}\end{figure}

\begin{figure}[t!]\begin{center}
\includegraphics[width=\columnwidth]{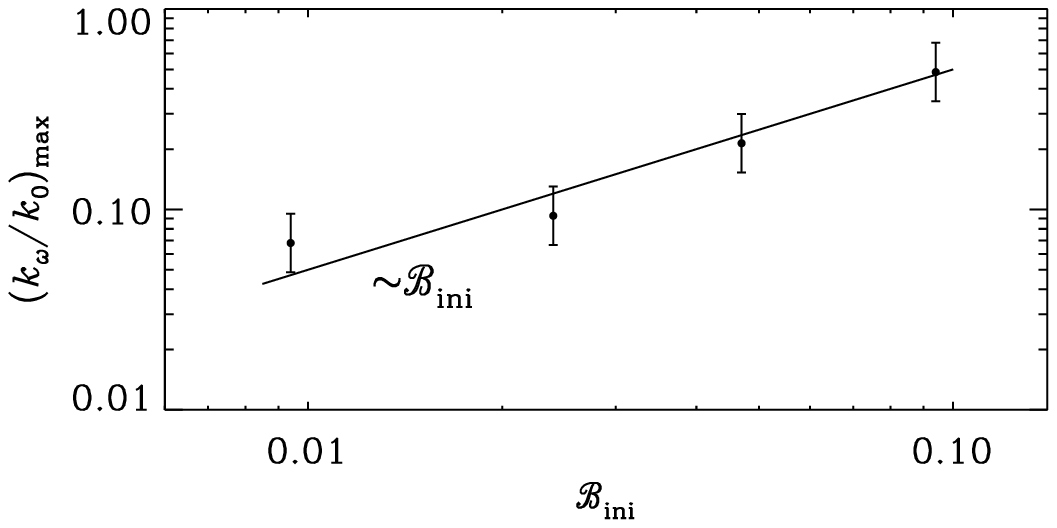}
\end{center}\caption{
Dependence of the maximum of $k_\omega/k_0$ on $\mathcal{B}_\mathrm{ini}$
for hydromagnetic runs with $\qzeta=1$,
$\Rm=900$, and different magnetic field strengths.
The straight line indicates a linear relationship.
Runs~35--38. 
}\label{pres_mag_assist}\end{figure}

In all our runs, $k_\omega/k_0$ reaches a maximum at some point.
For runs 15--18, we see that $(k_\omega/k_0)_{\max}$ increases
with increasing values of $\mathcal{B}_\mathrm{ini}$; see
\Fig{pcomp_qirro2_hydro_B}.
\Fig{pres_mag_assist} shows that this increase is linear and
not quadratic, which means that the vorticity is magnetically
driven rather than due to magnetically assisted growth; see
\cite{BS25} for details on this distinction.
As seen from \Tab{Tsummary}, the magnetic field decays for these runs,
so there is no dynamo action.
Due to the presence of the $a(t)$ factor in $\dot{\oo}_\mathrm{mag}$,
we expect the magnetic effect to diminish in collapse simulations with
a small value of $\kffN$.

\subsection{Spectral evolution}

\begin{figure}[t!]\begin{center}
\includegraphics[width=.94\columnwidth]{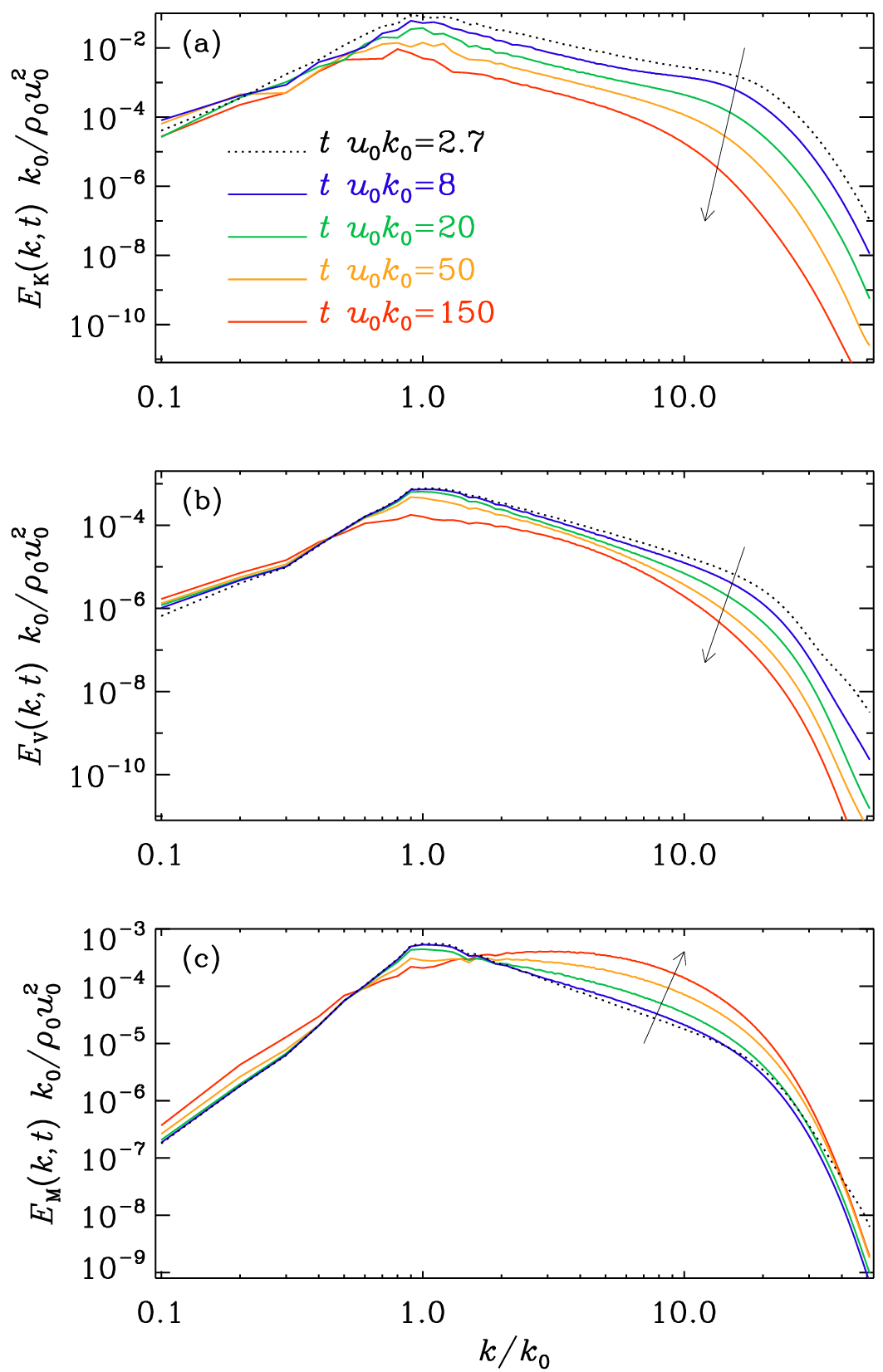}
\end{center}\caption{
Evolution of $\EK(k,t)$, $\EV(k,t)$, and $\EM(k,t)$ for Run~34.
The arrows indicate the sense of time.
The first time is shown as dotted lines to distinguish it better from
the next one, for which $\EM(k)$ is still very similar.
}\label{pspec_select_I096_1em3_1024c_eta2em6}\end{figure}

In \Fig{pspec_select_I096_1em3_1024c_eta2em6}, we show the evolution
of $\EK(k,t)$, $\EV(k,t)$, and $\EM(k,t)$ for Run~34.
This is our run with the largest magnetic Reynolds number ($\Rm=4500$)
and has only 4\% vorticity ($\qzeta=0.96$), but shows clear dynamo action.
The evolution of $k_\omega/k_0$ and $\calBrms$ was
shown in \Fig{pcomp_qirro3}.

\begin{figure}[t!]\begin{center}
\includegraphics[width=.94\columnwidth]{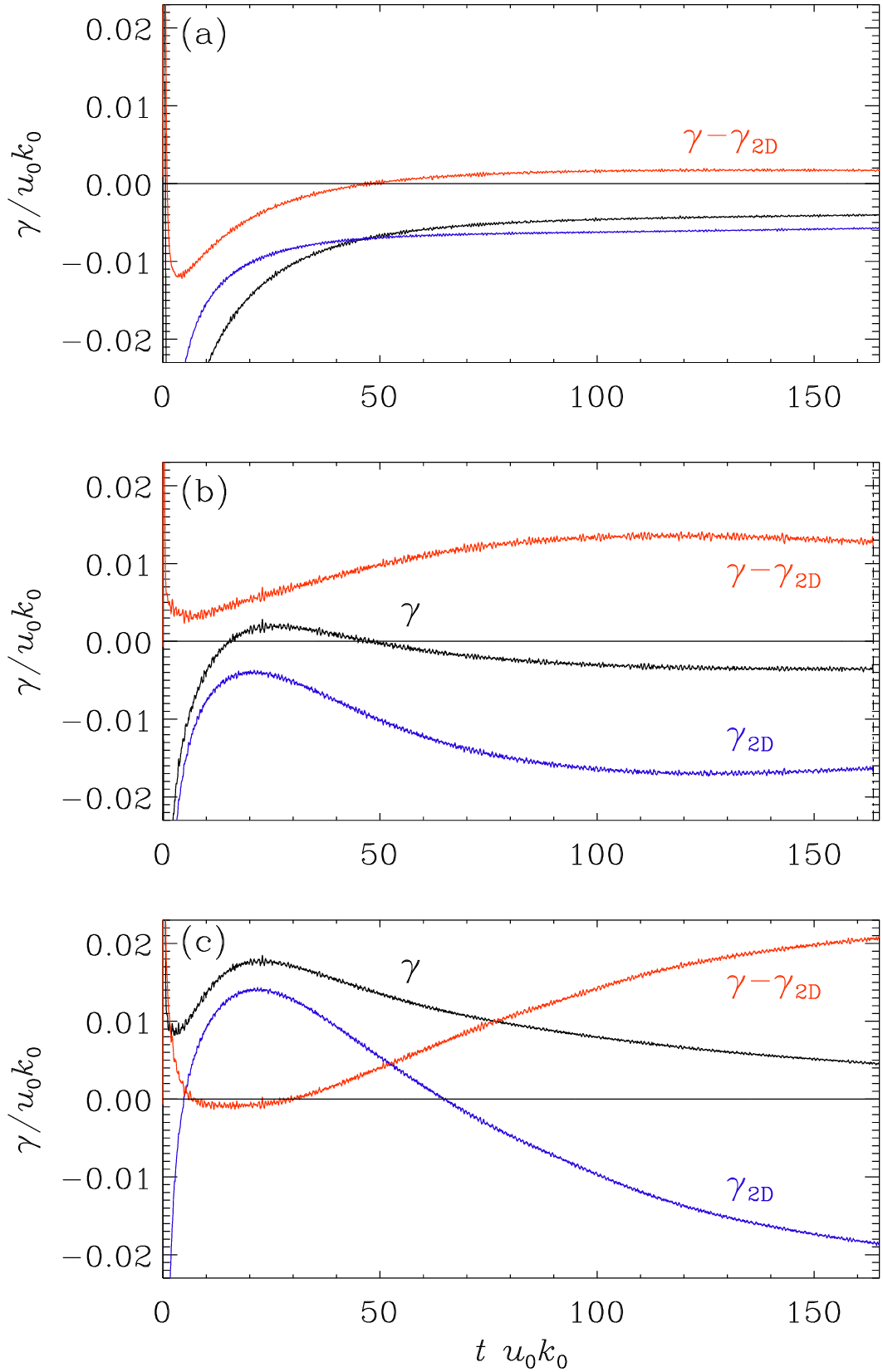}
\end{center}\caption{
Evolution of the pseudo growth rate $\gamma$ (black lines), with
contributions from $\gamma_{\rm 2D}$ (blue lines)
and the residual $\gamma-\gamma_{\rm 2D}$ (red lines),
for Runs~23 (a), 32 (b), and 34 (c). 
}\label{pWLcontri_comp}\end{figure}

\begin{figure*}[t!]\begin{center}
\includegraphics[width=.97\textwidth]{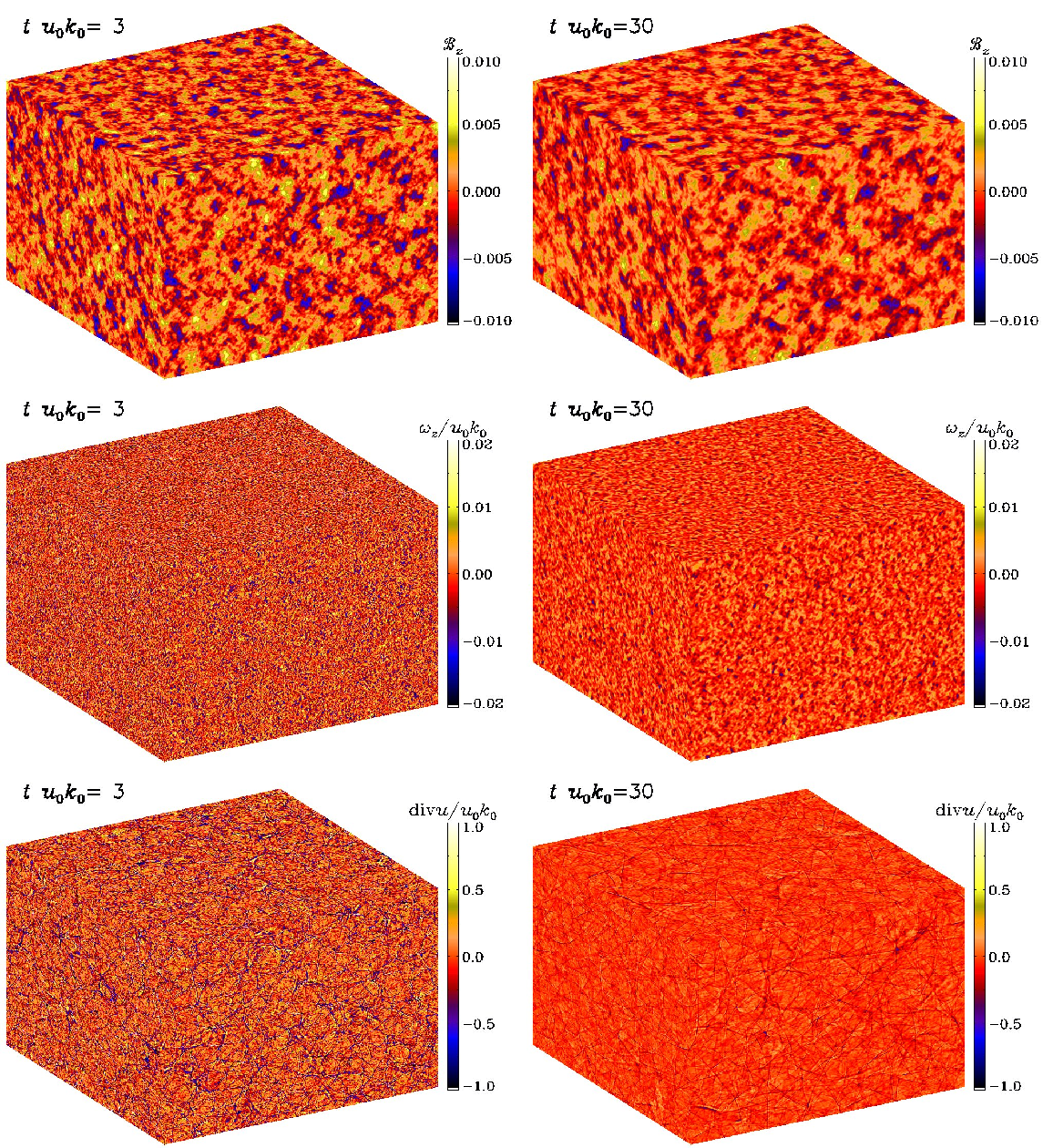}
\end{center}\caption{
Visualizations of $\mathcal{B}_z$, $\omega_z/u_0 k_0$, and
$\nab\cdot\uu/u_0 k_0$ for Run~37 at early and late times.
Note that the domain is cubic, but the images have been stretched in
the horizontal direction to take advantage of the full page size.
}\label{ABC}\end{figure*}

We see that both $\EK(k,t)$ and $\EV(k,t)$ decay, while $\EM(k,t)$
increases both at large and small wavenumbers.
Overall, $\EV(k)$ is almost a hundred times smaller than $\EK(k,t)$, but,
similarly to $\EM(k,t)$, $\EV(k)$ also shows a small temporal increase
at small values of $k$.
This is suggestive of magnetic vorticity production via an inverse
cascade.
Also, although $\EV(k,t)$ decays in the inertial range, it bulges at
$k/k_0\approx4$, which appears to be a direct consequence of magnetic
driving.

As already demonstrated in \cite{BN22}, the collapse dynamics do not
affect the magnetic energy spectra significantly.
At length scales above the Jeans length, the collapse does lead to
a growth of the compressive part of the kinetic energy spectra and
even a growth of magnetic energy, but this is associated with the
compression itself and is not a consequence of a dynamo; see Figure~9(b)
of \cite{BN22}.

\subsection{Instantaneous growth rate}

For the magnetic energy to grow,
the induction term $\uu\times\BB$
in \Eq{dAdt} has to overcome the dissipation term.
This is also true in the unsteady case and can therefore be used
to characterize dynamo action in a collapse simulation.
In the evolution equation for the mean magnetic energy density,
$\EEM(t)\equiv\bra{\BB^2/2\mu_0}$, the term
\begin{equation}
\bra{\JJ\cdot(\uu\times\BB)}\equiv-W_\mathrm{L}
\label{minusWL}
\end{equation}
has to exceed the Joule dissipation, $Q_{\rm M}=\bra{\mu_0\eta\JJ^2}$.
The instantaneous growth rate of magnetic energy can then be
written as $\gamma=(-W_{\rm L}-Q_{\rm M})/\EEM$.
The first term, which can also be written as
$W_{\rm L}=\bra{\uu\cdot(\JJ\times\BB)}$, is the work
done by the Lorentz force.
When it is negative, kinetic energy is used to drive magnetic energy;
see \Eq{minusWL}.

\cite{BN22} made use of the fact that in two dimensions (2D), when no action
is possible, \Eq{dAdt} can be written as an advection--diffusion equation,
i.e., $\DD A/\DD t=\eta\nabla^2A$, where $A$ is the component of $\AAA$
that is normal to the 2D plane.
This motivated them to decompose $W_{\rm L}$ by expanding
$\BB=\nab\times\AAA$ to get
\begin{equation}
-\bra{\JJ\cdot(\uu\times\BB)}=\bra{J_i u_j (A_{i,j}-A_{j,i}}
\equiv W_{\rm L}^{\rm2D}+W_{\rm L}^{\rm3D}.
\end{equation}
Here, the first term is related to the advection term.
The second term, $W_{\rm L}^{\rm 3D}=-\bra{J_i u_j A_{j,i}}$, vanishes
in 2D.
Thus, they identified $W_{\rm L}^{\rm3D}$ with a contribution that
characterizes the 3D nature of the system and used it as a proxy
for dynamo action when it is large enough.
They thus defined
\begin{equation}
\gamma_{\rm2D}=-(W_{\rm L}^{\rm2D}+Q_{\rm M})/\EEM,\quad
\gamma_{\rm3D}=-W_{\rm L}^{\rm3D}/\EEM,
\end{equation}
so that $\gamma_{\rm2D}+\gamma_{\rm3D}=\gamma$.

In \Fig{pWLcontri_comp}, we plot the time dependences of $\gamma$,
$\gamma_{\rm2D}$, and $\gamma_{\rm3D}=\gamma-\gamma_{\rm2D}$ for Runs~23
(no dynamo, because $k_\omega$ is too small), 32 (weak dynamo), and 34
(strong dynamo, $\Rm$ is the largest).  We see that $\gamma_{\rm2D}$
is always negative, except during an early phase for Run~34, which can
be associated with strong 2D tangling of the initial magnetic field.
When $\gamma_{\rm3D}$ is added to $\gamma_{\rm2D}$, the resulting
instantaneous growth rate is positive during the early part of the
evolution of Run~32 and during the entire evolution of Run~34.

\begin{figure*}[t!]\begin{center}
\includegraphics[width=\textwidth]{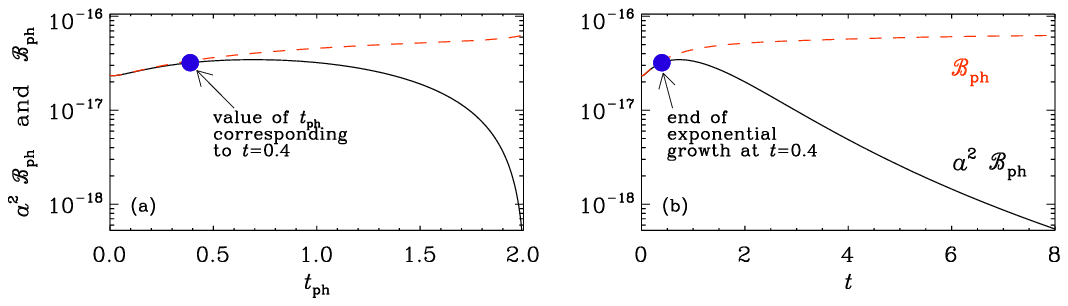}
\end{center}\caption{
Physical magnetic field $\mathcal{B}_\mathrm{ph}$ (dashed red lines)
and its comoving counterpart $a^2 \mathcal{B}_\mathrm{ph}$ (black lines)
versus physical time (a) and conformal time (b) for Run~B from
\cite{BN22} and Run~39 of the present paper.
}\label{preverse}\end{figure*}

Our considerations above have shown that the use of $\gamma_{\rm2D}$ and
$\gamma_{\rm3D}$ does indeed provide a meaningful tool to assess dynamo
action in unsteady environments in general, and in collapse simulations
in particular.
Nevertheless, we regard the direct demonstration of exponential growth
in supercomoving coordinates in \Sec{GrowthConformal} as even more
convincing evidence for dynamo action.

\subsection{Visualizations}

In \Fig{ABC}, we present visualizations of $\mathcal{B}_z$,
$\omega_z/u_0 k_0$, and $\nab\cdot\uu/u_0 k_0$ for Run~37 at early and late times.
There is no significance in our having chosen the $z$ component of $\BB$ and $\oo$;
all three components are statistically equivalent.

The magnetic field appears to preserve its initial length scale
corresponding to $k=k_0$, and only the field strength becomes weaker
with time.
By contrast, the vorticity quickly develops small-scale patches that
then grow to larger-scale patches at later times.
Note also that the magnitude of $\omega_z/u_0 k_0$ (about 0.01) is
comparable to that of $\mathcal{B}_z$.
This is reminiscent of the findings of \cite{Kahn+12}, who reported a
quantitative agreement between the spectra of vorticity and magnetic field.

For the velocity divergence, there is a much larger decrease from
the time $t u_0 k_0=3$ to $t u_0 k_0=30$.
As stated above, the compressive part of the velocity field, which is
reflected in the values and the appearance of $\nab\cdot\uu$, decreases
more strongly with time than the vortical part, as reflected through
the vorticity.
We also see that, although the initial scales are rather small, they
still seem to be sufficiently well resolved.

\section{Comparison with previous work}
\label{HomogeneousCollapse}

In our earlier paper \citep{BN22}, we simulated gravitational collapse
using numerical simulations of decaying turbulence in a Jeans-unstable
domain at a resolution of $2048^2$ mesh points.
We only found a weak increase in the magnetic field with time.
Given the knowledge of the collapse time from the simulations,
i.e., the time when the singularity was reached, we can
replace the pressureless freefall time by the actual collapse time and
express the evolution of the rms magnetic field in comoving coordinates.
This allows us to see whether the growth in the old simulations
is close to exponential in comoving coordinates
during any time interval.

The result is shown in \Fig{preverse}, where we computed the conformal
time and scale factor numerically based on \Eq{ascale}.
Here we used the empirical value of $\tff\approx2.016/\cs k_1$,
which yields $\kff\approx0.78\,\cs k_0$, and thus, since $u_0/\cs=0.19$
and $k_0/k_1=10$, we have $\kffN\approx0.4$; see
\Tab{Tsummary}, where it is called Run~39.
The physical values of the magnetic field computed by \cite{BN22} are denoted by
$\mathcal{B}_\mathrm{ph}$.
We also plot the comoving values $a^2 \mathcal{B}_\mathrm{ph}$ both
versus physical and conformal time.
Here, the $a(t)$ and the conformal time have been computed from \Eqs{ascale}{tph}.
Although there is a steady increase of $\calBrms$, \Fig{preverse}(b) shows
that the comoving magnetic field does not follow an exponential growth
in conformal time, except for a very early time interval $0<t u_0 k_0 \la 0.4$.

To understand why the exponential phase is so short in this run, we
compare its parameters with those of the other runs presented in this
paper; see \Tab{Tsummary}.
The closest match is with Run~1.
We see immediately that the main problem with Run~39 is the small value
of the magnetic Reynolds number, which is 10 times smaller than that
of Run~1.
In spite of the high resolution of Run~39, the value of $\Rm$ could
not have been chosen larger because of the strong compression and large
gradients suffered by the collapsing regions toward the end of the run.
This highlights the main advantage of choosing supercomoving coordinates
for collapse simulations.

\section{Conclusions}

In this work we approached the problem of dynamo action during
gravitational collapse by employing supercomoving coordinates.
This is a significant change of paradigm with respect to previous
simulation work \citep[e.g.,][]{Sur+10,Sur+12,Federrath+11b,BN22}, which
was limited by the shrinking dynamical range during the collapse;
see \Sec{HomogeneousCollapse}.
In supercomoving coordinates, we can look for exponential growth of the
magnetic field, which is a clear signature of a dynamo.
This allows us to surpass the other obstacle faced by previous work,
which is characterizing dynamos in unsteady flows.

When describing gravitational collapse in supercomoving coordinates,
the governing equations of magnetohydrodynamics are similar to the
original ones, except that now the scale factor appears in front
of the Lorentz force.
This reduces the effective Lorentz force, because $a(t)$ becomes
progressively smaller with time.
Therefore, in the limit of very short collapse times or large values
of $\kff$, the evolution approaches essentially the kinematic evolution.
This, however, does not mean unlimited continual growth, because the rms
value of the turbulent intensity is declining.

As shown previously \citep{Bra+19}, decaying turbulence leads to an
episode of exponential growth if the magnetic Reynolds number is large
enough.
The larger it is, the longer is the episode of exponential growth.
This is essentially the result of a competition against the decay of
turbulence, which lowers the instantaneous value of the magnetic
Reynolds number as time goes on.
The gravitational collapse changes this picture only little if we view
the decay in supercomoving coordinates, because the collapse only affects
the nonlinear dynamics, and this nonlinearity gets weaker with time.

\cite{Irshad+25} considered forced turbulence, as opposed to our study
of decaying turbulence.
Therefore, in their models, the magnetic field could always be
sustained, but the source of the driving remains unclear.
The superexponential growth that they reported, however, was still
recovered in our decay simulations, unless the freefall time is longer
than the turnover time of the turbulence.
In that case, the growth is actually subexponential,
but this is primarily a consequence of the decay of the turbulence.

Our present work has also shown that even very small amounts of vorticity
can be sufficient to facilitate dynamo action.
In particular, we find that the vorticity can grow in concert with the
magnetic field.
However, the magnetic vorticity production will decline in simulations
with small values of $\kffN$.

Earlier work on turbulent collapse and dynamo action has suggested that
the collapse drives turbulence and enhances it \citep{Sur+12,
XL20, Hennebelle21}.
Our work casts doubt on this interpretation due to two aspects.
First, the collapse dynamics reduces the effective nonlinearity, resulting
in stronger apparent field amplification by the turbulence, and second,
there can be generation of vorticity both from viscosity and from the
magnetic field itself, but this is not directly related to the collapse.
It should therefore be checked whether these two factors could have
contributed to the earlier findings of collapse-driven turbulence.
In this context, the fact that we do not solve the Poisson equation for self-gravity but treat the collapse as a homogeneous flow
through the change of coordinates could be a difference worth investigating.

As explained in \Sec{HomogeneousCollapse}, the transformation to
supercomoving coordinates may also help analyze existing simulations
in physical coordinates.
We argue that for homogeneous collapse simulations that do not utilize
supercomoving coordinates, it is still useful to express such results
in terms of comoving quantities and conformal time, because they
might display exponential magnetic field growth that would be perhaps
the strongest indication of dynamo action so far.

Our work has applications not just to interstellar clouds and primordial
star formation \citep[e.g.,][]{Schleicher2009,hirano2022,Sharda2020},
but also to larger cosmological scales.
Our results show that small amounts of vorticity might suffice to produce
dynamo action even in decaying turbulence, which, we argue, is also
relevant to gravitational collapse.
This consideration is important for understanding magnetism in protohalos
before the first stars form and their feedback drives sufficient
turbulence for dynamo action \citep[e.g.,][]{Schleicher2010}.

Finally, our findings indicate that earlier simulations, including our
own high-resolution simulations at $2048^3$ mesh points, may still have
had insufficient resolution to follow the collapse and should be revisited
using more idealized settings that allow the usage of a comoving frame.

\begin{acknowledgments}
We thank Fabio Del Sordo for helpful discussions on vorticity generation,
Pallavi Bhat for suggesting an explanation for the subexponential
growth for $\kffN\ll1$, and the referee for a detailed assessment of our work.
We also acknowledge inspiring discussions with the participants of
the program on ``Turbulence in Astrophysical Environments'' at the Kavli
Institute for Theoretical Physics in Santa Barbara.
This research was supported in part by the Swedish Research Council
(Vetenskapsr{\aa}det) under grant No.\ 2019-04234, the National Science
Foundation under grants No.\ NSF PHY-2309135, AST-2307698, AST-2408411,
and NASA Award 80NSSC22K0825.
E.N.\ acknowledges funding from the Italian Ministry for Universities and Research
(MUR) through the ``Young Researchers'' funding call (Project MSCA 000074).
We acknowledge the allocation of computing resources provided by the
Swedish National Allocations Committee at the Center for
Parallel Computers at the Royal Institute of Technology in Stockholm.

\vspace{2mm}\noindent
{\em Software and Data Availability.}
The source code used for the simulations of this study,
the {\sc Pencil Code} \citep{PC}, is freely available on
\url{https://github.com/pencil-code}.
The simulation setups and corresponding input
and reduced output data are freely available on
\dataset[http://doi.org/10.5281/zenodo.15693287]{http://doi.org/10.5281/zenodo.15693287}.
\end{acknowledgments}

\bibliographystyle{aasjournal}
\bibliography{ref}

\begin{thebibliography}{}
\expandafter\ifx\csname natexlab\endcsname\relax\def\natexlab#1{#1}\fi
\providecommand{\url}[1]{\href{#1}{#1}}
\providecommand{\dodoi}[1]{doi:~\href{http://doi.org/#1}{\nolinkurl{#1}}}
\providecommand{\doeprint}[1]{\href{http://ascl.net/#1}{\nolinkurl{http://ascl.net/#1}}}
\providecommand{\doarXiv}[1]{\href{https://arxiv.org/abs/#1}{\nolinkurl{https://arxiv.org/abs/#1}}}

\bibitem[{{Achikanath Chirakkara} {et~al.}(2021){Achikanath Chirakkara},
  {Federrath}, {Trivedi}, \& {Banerjee}}]{AchikanathChirakkara+21}
{Achikanath Chirakkara}, R., {Federrath}, C., {Trivedi}, P., \& {Banerjee}, R.
  2021, \prl, 126, 091103, \dodoi{10.1103/PhysRevLett.126.091103}

\bibitem[{{Batchelor}(1950)}]{Batchelor50}
{Batchelor}, G.~K. 1950, RSPSA, 201, 405, \dodoi{10.1098/rspa.1950.0069}

\bibitem[{{Beck} {et~al.}(1994){Beck}, {Poezd}, {Shukurov}, \&
  {Sokoloff}}]{Beck+94}
{Beck}, R., {Poezd}, A.~D., {Shukurov}, A., \& {Sokoloff}, D.~D. 1994, \aap,
  289, 94

\bibitem[{{Brandenburg} {et~al.}(2019){Brandenburg}, {Kahniashvili}, {Mandal},
  {Pol}, {Tevzadze}, \& {Vachaspati}}]{Bra+19}
{Brandenburg}, A., {Kahniashvili}, T., {Mandal}, S., {et~al.} 2019, PhRvF, 4,
  024608, \dodoi{10.1103/PhysRevFluids.4.024608}

\bibitem[{{Brandenburg} \& {Ntormousi}(2022)}]{BN22}
{Brandenburg}, A., \& {Ntormousi}, E. 2022, \mnras, 513, 2136,
  \dodoi{10.1093/mnras/stac982}

\bibitem[{{Brandenburg} \& {Ntormousi}(2023)}]{BN23}
---. 2023, \araa, 61, 561, \dodoi{10.1146/annurev-astro-071221-052807}

\bibitem[{{Brandenburg} \& {Scannapieco}(2025)}]{BS25}
{Brandenburg}, A., \& {Scannapieco}, E. 2025, \apj, 983, 105,
  \dodoi{10.3847/1538-4357/adbe38}

\bibitem[{{Chen} {et~al.}(2024){Chen}, {Lopez-Rodriguez}, {Ivison}, {Geach},
  {Dye}, {Liu}, \& {Bendo}}]{Chen2024}
{Chen}, J., {Lopez-Rodriguez}, E., {Ivison}, R.~J., {et~al.} 2024, \aap, 692,
  A34, \dodoi{10.1051/0004-6361/202450969}

\bibitem[{{Cowling}(1933)}]{Cow33}
{Cowling}, T.~G. 1933, \mnras, 94, 39, \dodoi{10.1093/mnras/94.1.39}

\bibitem[{{Del Sordo} \& {Brandenburg}(2011)}]{DSB11}
{Del Sordo}, F., \& {Brandenburg}, A. 2011, \aap, 528, A145,
  \dodoi{10.1051/0004-6361/201015661}

\bibitem[{{Elias-L{\'o}pez} {et~al.}(2023){Elias-L{\'o}pez}, {Del Sordo}, \&
  {Vigan{\`o}}}]{Elias-Lopez+23}
{Elias-L{\'o}pez}, A., {Del Sordo}, F., \& {Vigan{\`o}}, D. 2023, \aap, 677,
  A46, \dodoi{10.1051/0004-6361/202346696}

\bibitem[{{Elias-L{\'o}pez} {et~al.}(2024){Elias-L{\'o}pez}, {Del Sordo}, \&
  {Vigan{\`o}}}]{Elias-Lopez+24}
---. 2024, \aap, 690, A77, \dodoi{10.1051/0004-6361/202450398}

\bibitem[{{Federrath} {et~al.}(2011{\natexlab{a}}){Federrath}, {Chabrier},
  {Schober}, {Banerjee}, {Klessen}, \& {Schleicher}}]{Federrath+11}
{Federrath}, C., {Chabrier}, G., {Schober}, J., {et~al.} 2011{\natexlab{a}},
  \prl, 107, 114504, \dodoi{10.1103/PhysRevLett.107.114504}

\bibitem[{{Federrath} {et~al.}(2011{\natexlab{b}}){Federrath}, {Sur},
  {Schleicher}, {Banerjee}, \& {Klessen}}]{Federrath+11b}
{Federrath}, C., {Sur}, S., {Schleicher}, D. R.~G., {Banerjee}, R., \&
  {Klessen}, R.~S. 2011{\natexlab{b}}, \apj, 731, 62,
  \dodoi{10.1088/0004-637X/731/1/62}

\bibitem[{{Geach} {et~al.}(2023){Geach}, {Lopez-Rodriguez}, {Doherty}, {Chen},
  {Ivison}, {Bendo}, {Dye}, \& {Coppin}}]{Geach2023}
{Geach}, J.~E., {Lopez-Rodriguez}, E., {Doherty}, M.~J., {et~al.} 2023, \nat,
  621, 483, \dodoi{10.1038/s41586-023-06346-4}

\bibitem[{{Gilbert} {et~al.}(1988){Gilbert}, {Frisch}, \&
  {Pouquet}}]{Gilbert+88}
{Gilbert}, A.~D., {Frisch}, U., \& {Pouquet}, A. 1988, GApFD, 42, 151,
  \dodoi{10.1080/03091928808208861}

\bibitem[{{Haugen} {et~al.}(2004){Haugen}, {Brandenburg}, \& {Mee}}]{HBM04}
{Haugen}, N. E.~L., {Brandenburg}, A., \& {Mee}, A.~J. 2004, \mnras, 353, 947,
  \dodoi{10.1111/j.1365-2966.2004.08127.x}

\bibitem[{{Hennebelle}(2021)}]{Hennebelle21}
{Hennebelle}, P. 2021, \aap, 655, A3, \dodoi{10.1051/0004-6361/202141650}

\bibitem[{{Hide} \& {Palmer}(1982)}]{Hide+Palmer82}
{Hide}, R., \& {Palmer}, T.~N. 1982, GApFD, 19, 301,
  \dodoi{10.1080/03091928208208961}

\bibitem[{{Hirano} \& {Machida}(2022)}]{hirano2022}
{Hirano}, S., \& {Machida}, M.~N. 2022, \apjl, 935, L16,
  \dodoi{10.3847/2041-8213/ac85e0}

\bibitem[{{Irshad P} {et~al.}(2025){Irshad P}, {Bhat}, {Subramanian}, \&
  {Shukurov}}]{Irshad+25}
{Irshad P}, M., {Bhat}, P., {Subramanian}, K., \& {Shukurov}, A. 2025, arXiv
  e-prints, arXiv:2503.19131, \dodoi{10.48550/arXiv.2503.19131}

\bibitem[{{Kahniashvili} {et~al.}(2012){Kahniashvili}, {Brandenburg},
  {Campanelli}, {Ratra}, \& {Tevzadze}}]{Kahn+12}
{Kahniashvili}, T., {Brandenburg}, A., {Campanelli}, L., {Ratra}, B., \&
  {Tevzadze}, A.~G. 2012, \prd, 86, 103005, \dodoi{10.1103/PhysRevD.86.103005}

\bibitem[{{Kazantsev}(1968)}]{Kaz68}
{Kazantsev}, A.~P. 1968, JETP, 26, 1031

\bibitem[{{Kazantsev} {et~al.}(1985){Kazantsev}, {Ruzmaikin}, \&
  {Sokolov}}]{Kazantsev+85}
{Kazantsev}, A.~P., {Ruzmaikin}, A.~A., \& {Sokolov}, D.~D. 1985, ZhETF, 61,
  285

\bibitem[{{Kulsrud} \& {Anderson}(1992)}]{Kulsrud+Anderson92}
{Kulsrud}, R.~M., \& {Anderson}, S.~W. 1992, \apj, 396, 606,
  \dodoi{10.1086/171743}

\bibitem[{{Martel} \& {Shapiro}(1998)}]{MS98}
{Martel}, H., \& {Shapiro}, P.~R. 1998, \mnras, 297, 467,
  \dodoi{10.1046/j.1365-8711.1998.01497.x}

\bibitem[{{Martins Afonso} {et~al.}(2019){Martins Afonso}, {Mitra}, \&
  {Vincenzi}}]{Afonso+19}
{Martins Afonso}, M., {Mitra}, D., \& {Vincenzi}, D. 2019, RSPSA, 475,
  20180591, \dodoi{10.1098/rspa.2018.0591}

\bibitem[{{Mee} \& {Brandenburg}(2006)}]{MB06}
{Mee}, A.~J., \& {Brandenburg}, A. 2006, \mnras, 370, 415,
  \dodoi{10.1111/j.1365-2966.2006.10476.x}

\bibitem[{{Meneguzzi} {et~al.}(1981){Meneguzzi}, {Frisch}, \&
  {Pouquet}}]{Meneguzzi+81}
{Meneguzzi}, M., {Frisch}, U., \& {Pouquet}, A. 1981, \prl, 47, 1060,
  \dodoi{10.1103/PhysRevLett.47.1060}

\bibitem[{{Moffatt} \& {Proctor}(1985)}]{MP85}
{Moffatt}, H.~K., \& {Proctor}, M.~R.~E. 1985, \jfm, 154, 493,
  \dodoi{10.1017/S002211208500163X}

\bibitem[{{Parker}(1955)}]{Par55}
{Parker}, E.~N. 1955, \apj, 122, 293, \dodoi{10.1086/146087}

\bibitem[{{Parker}(1971)}]{Par71}
---. 1971, \apj, 163, 255, \dodoi{10.1086/150765}

\bibitem[{{Pattle} {et~al.}(2023){Pattle}, {Fissel}, {Tahani}, {Liu}, \&
  {Ntormousi}}]{Pattle23}
{Pattle}, K., {Fissel}, L., {Tahani}, M., {Liu}, T., \& {Ntormousi}, E. 2023,
  in Astronomical Society of the Pacific Conference Series, Vol. 534,
  Protostars and Planets VII, ed. S.~{Inutsuka}, Y.~{Aikawa}, T.~{Muto},
  K.~{Tomida}, \& M.~{Tamura}, 193, \dodoi{10.48550/arXiv.2203.11179}

\bibitem[{{Pencil Code Collaboration} {et~al.}(2021){Pencil Code
  Collaboration}, {Brandenburg}, {Johansen}, {Bourdin}, {Dobler}, {Lyra},
  {Rheinhardt}, {Bingert}, {Haugen}, {Mee}, {Gent}, {Babkovskaia}, {Yang},
  {Heinemann}, {Dintrans}, {Mitra}, {Candelaresi}, {Warnecke},
  {K{\"a}pyl{\"a}}, {Schreiber}, {Chatterjee}, {K{\"a}pyl{\"a}}, {Li},
  {Kr{\"u}ger}, {Aarnes}, {Sarson}, {Oishi}, {Schober}, {Plasson}, {Sandin},
  {Karchniwy}, {Rodrigues}, {Hubbard}, {Guerrero}, {Snodin}, {Losada},
  {Pekkil{\"a}}, \& {Qian}}]{PC}
{Pencil Code Collaboration}, {Brandenburg}, A., {Johansen}, A., {et~al.} 2021,
  JOSS, 6, 2807, \dodoi{10.21105/joss.02807}

\bibitem[{{Schleicher} {et~al.}(2010){Schleicher}, {Banerjee}, {Sur},
  {Arshakian}, {Klessen}, {Beck}, \& {Spaans}}]{Schleicher2010}
{Schleicher}, D.~R.~G., {Banerjee}, R., {Sur}, S., {et~al.} 2010, \aap, 522,
  A115, \dodoi{10.1051/0004-6361/201015184}

\bibitem[{{Schleicher} {et~al.}(2009){Schleicher}, {Galli}, {Glover},
  {Banerjee}, {Palla}, {Schneider}, \& {Klessen}}]{Schleicher2009}
{Schleicher}, D. R.~G., {Galli}, D., {Glover}, S. C.~O., {et~al.} 2009, \apj,
  703, 1096, \dodoi{10.1088/0004-637X/703/1/1096}

\bibitem[{{Schober} {et~al.}(2012){Schober}, {Schleicher}, {Federrath},
  {Glover}, {Klessen}, \& {Banerjee}}]{Schober2012}
{Schober}, J., {Schleicher}, D., {Federrath}, C., {et~al.} 2012, \apj, 754, 99,
  \dodoi{10.1088/0004-637X/754/2/99}

\bibitem[{{Shandarin}(1980)}]{Shandarin80}
{Shandarin}, S.~F. 1980, Astrophysics, 16, 439, \dodoi{10.1007/BF01005530}

\bibitem[{{Sharda} {et~al.}(2020){Sharda}, {Federrath}, \&
  {Krumholz}}]{Sharda2020}
{Sharda}, P., {Federrath}, C., \& {Krumholz}, M.~R. 2020, \mnras, 497, 336,
  \dodoi{10.1093/mnras/staa1926}

\bibitem[{{Steenbeck} {et~al.}(1966){Steenbeck}, {Krause}, \&
  {R{\"a}dler}}]{SKR66}
{Steenbeck}, M., {Krause}, F., \& {R{\"a}dler}, K.~H. 1966, Zeitschrift
  Naturforschung Teil A, 21, 369, \dodoi{10.1515/zna-1966-0401}

\bibitem[{{Subramanian} \& {Brandenburg}(2014)}]{SB14}
{Subramanian}, K., \& {Brandenburg}, A. 2014, \mnras, 445, 2930,
  \dodoi{10.1093/mnras/stu1954}

\bibitem[{{Sur} {et~al.}(2012){Sur}, {Federrath}, {Schleicher}, {Banerjee}, \&
  {Klessen}}]{Sur+12}
{Sur}, S., {Federrath}, C., {Schleicher}, D. R.~G., {Banerjee}, R., \&
  {Klessen}, R.~S. 2012, \mnras, 423, 3148,
  \dodoi{10.1111/j.1365-2966.2012.21100.x}

\bibitem[{{Sur} {et~al.}(2010){Sur}, {Schleicher}, {Banerjee}, {Federrath}, \&
  {Klessen}}]{Sur+10}
{Sur}, S., {Schleicher}, D.~R.~G., {Banerjee}, R., {Federrath}, C., \&
  {Klessen}, R.~S. 2010, \apjl, 721, L134, \dodoi{10.1088/2041-8205/721/2/L134}

\bibitem[{{Vainshtein} \& {Ruzmaikin}(1971)}]{VR71}
{Vainshtein}, S.~I., \& {Ruzmaikin}, A.~A. 1971, \azh, 48, 902

\bibitem[{{Xu} \& {Lazarian}(2020)}]{XL20}
{Xu}, S., \& {Lazarian}, A. 2020, \apj, 899, 115,
  \dodoi{10.3847/1538-4357/aba7ba}

\end{thebibliography}

\end{document}